\documentclass[amsmath,amssymb,aps,twocolumn,superscriptaddress]{revtex4-1}

\usepackage[total={7.0in,9.5in}, top=1.0in, includefoot]{geometry}

\usepackage[usenames,dvipsnames,svgnames,table]{xcolor}
\usepackage{graphicx}

\newcommand{\sam}[1]{{\color{black} #1}}
\newcommand{\dan}[1]{{\color{black} #1}}
\newcommand{\aaron}[1]{{\color{black} #1}}
\newcommand{\allie}[1]{{\color{black} #1}}

\usepackage{hyperref}
\hypersetup{
    hypertexnames=false,
    bookmarks=true,         
    unicode=false,          
    pdftoolbar=true,        
    pdfmenubar=true,        
    pdffitwindow=false,     
    pdfstartview={FitH},    
    pdfcreator={},   
    pdfproducer={}, 
    pdfkeywords={} {} {}, 
    pdfnewwindow=true,      
    colorlinks=true,       
    linkcolor=red,          
    citecolor=Green,        
    filecolor=magenta,      
    urlcolor=cyan           
}

\newcommand{\figfolder}{}

\begin{document}

\author{\sam{Samuel F. Way}}
	\email{samuel.way@colorado.edu}
	\affiliation{Department of Computer Science, University of Colorado, Boulder, CO, USA}
\author{\allie{Allison C. Morgan}}
	\email{allison.morgan@colorado.edu}
	\affiliation{Department of Computer Science, University of Colorado, Boulder, CO, USA}
\author{\aaron{Aaron Clauset}}
	\email{aaron.clauset@colorado.edu}
	\affiliation{Department of Computer Science, University of Colorado, Boulder, CO, USA}
	\affiliation{BioFrontiers Institute, University of Colorado, Boulder, CO, USA}
	\affiliation{Santa Fe Institute, Santa Fe, NM, USA}
\author{\dan{Daniel B. Larremore}}
	\email{daniel.larremore@colorado.edu}
	\affiliation{Department of Computer Science, University of Colorado, Boulder, CO, USA}
	\affiliation{BioFrontiers Institute, University of Colorado, Boulder, CO, USA}
	\affiliation{Santa Fe Institute, Santa Fe, NM, USA}

\title{The misleading narrative of the canonical faculty productivity trajectory}

\begin{abstract}
A scientist may publish tens or hundreds of papers over a career, but these contributions are not evenly spaced in time. Sixty years of studies on career productivity patterns in a variety of fields suggest an intuitive and universal pattern: productivity tends to rise rapidly to an early peak and then gradually declines. Here, we test the universality of this conventional narrative by analyzing the structures of individual faculty productivity time series, constructed from over 200,000 publications and matched with hiring data for 2453 tenure-track faculty in all 205 Ph.D-granting computer science departments in the U.S. and Canada. Unlike prior studies, which considered only some faculty or some institutions, or lacked common career reference points, here we combine a large bibliographic dataset with comprehensive information on career transitions that covers an entire field of study. We show that the conventional narrative confidently describes only one fifth of faculty, regardless of department prestige or researcher gender, and the remaining four fifths of faculty exhibit a rich diversity of productivity patterns. To explain this diversity, we introduce a simple model of productivity trajectories, and explore correlations between its parameters and researcher covariates, showing that departmental prestige predicts overall individual productivity and the timing of the transition from first- to last-author publications. These results demonstrate the unpredictability of productivity over time, and open the door for new efforts to understand how environmental and individual factors shape scientific productivity.
\end{abstract}

\maketitle

\section*{Introduction}
\vspace{-0.15in}

Scholarly publications serve as the primary mode of communication through which scientific knowledge is developed, discussed, and disseminated. The amount that an individual researcher contributes to this dialogue---their scholarly productivity---thus serves as an important measure of the rate at which they contribute units of knowledge to the field, and this measure is known to influence the placement of graduates into faculty jobs \cite{way2016gender}, the likelihood of being granted tenure \cite{cole1967scientific,massy95}, and the ability to secure funding for future research \cite{stephan1996economics}. 

The trajectory of productivity over the course of a researcher's lifetime has been studied for at least 60 years, with the common observation being that a researcher's productivity rises rapidly to a peak and then slowly declines \cite{lehman1954men,dennis1956age,diamond1986life,horner1986relation,bayer1977career}, 
which has inspired the construction of mechanistic models with a similar profile \cite{bayer1977career,simonton1997creative,diamond1986life,levin1991research,rinaldi2000instabilities}.  
These models have included factors like cognitive decline with age, career age, finite supplies of human capital, knowledge advantages conferred by recent education, as well as skill deficits among the young, among others, and have been supported by the observation that individual productivity curves feature both long- and medium-term fluctuations \cite{rinaldi2000instabilities} and are not well described by even fourth-degree polynomial models \cite{bayer1977career}.
Indeed, every study we found to date proposes or confirms a ``rise and decline,'' ``curvilinear,'' or ``peak and tapering'' productivity trajectory, regardless of whether researchers are binned by chronological age \cite{lehman1954men,dennis1956age,cole1979age,diamond1986life,horner1986relation,simonton1997creative,levin1991research,rinaldi2000instabilities}, career age \cite{bayer1977career,simonton1997creative}, or (only for young researchers) years since first publication \cite{symonds2006gender}. 
The pattern may even extend to mentorship, supported by a finding that the prot\'eg\'es of early-career mathematicians tended to mentor more students, themselves, than prot\'eg\'es trained by those same faculty late in their careers \cite{malmgren2010role}. In fact, this conventional narrative of the life course is not restricted to academia, with similar trajectories observed in criminal behavior and artistic production in 1800s France \cite{quetelet1835homme} and even productivity of food acquisition by hunter-gatherers \cite{kaplan2000theory}.

While these past studies have firmly established that the conventional academic productivity narrative is equally descriptive across fields and time, their analyses are based on averages over hundreds or thousands of individuals \cite{lehman1954men,dennis1956age,cole1979age,diamond1986life,horner1986relation,simonton1997creative,levin1991research,bayer1977career,symonds2006gender,quetelet1835homme,kaplan2000theory,malmgren2010role}. This raises two crucial and previously unanswered questions: is this average trajectory representative of individual faculty, and how much diversity is hidden by a focus on a central tendency over a population? To answer these questions, we combine and study two comprehensive datasets that span forty years of productivity for nearly every tenure-track professor in a North American Ph.D.-granting computer science department. By introducing a simple mathematical description of the shape of a scientist's productivity over time, we map individuals' publication histories to a low-dimensional parameter space, revealing substantial diversity in the publication trends of individual faculty and showing that only a minority follow the conventional narrative of productivity. In fact, even among the conventional trajectories, individuals exhibit large fluctuations in their productivity around the average trend. Together, these results reveal that population averages provide a dramatically inaccurate picture of intellectual contributions over time, and that productivity patterns are both more diverse and less predictable than previously thought. These findings were preliminarily described in a recent review \cite{clauset2017data} which provides additional context for the results reported fully here.

Moreover, while we show that the distribution of productivity trajectories resists natural categorization, it is nevertheless possible to explore covariates that are associated with different regions of its parameter space. The literature on such associations has avoided detailed trajectories and instead focused on the complicated relationship between prestige, productivity, and hiring. Past studies have found that researchers trained at prestigious institutions are likely to remain productive \cite{caplow1958academic}, regardless of where they place as faculty \cite{crane1965scientists}. Other results link the prestige of the doctorate and the advisor to early-career productivity but not long-term productivity \cite{reskin1979academic}, which is at odds with others \cite{long1979entrance,chubin1981career} who found that early-career productivity predicts long-term productivity.
Disagreement about hiring exists as well, with multiple studies finding that doctoral prestige, and not productivity drives the initial placement of faculty \cite{zuckerman1970stratification,long1978productivity} 
while recent work based on comprehensive data in multiple fields suggests that prestige alone is insufficient to fully explain faculty placement \cite{clauset2015systematic,way2016gender}. This, too, is complicated by hypotheses of mutual causality, where departments both select for and facilitate high productivity \cite{allison1990departmental}.
Unfortunately, while such studies shed light on a complicated system, they tend to restrict their analyses to unusual scientists, such as Nobel laureates or faculty at elite departments, rather than typical researchers.  In contrast, the data analyzed here are comprehensive, covering faculty across the prestige hierarchy, which enables us to move beyond total productivity to study publication trajectories in light of prestige, hiring, and past productivity alike.

This study exploits and combines two large datasets related to faculty productivity. The first one is a comprehensive, hand-curated collection of education and academic appointment histories for tenure-track and tenured computer science faculty \cite{clauset2015systematic}. This dataset spans all 205 departmental or school-level academic units on the Computing Research Association's Forsythe List of Ph.D.-granting departments in computing-related disciplines in the United States and Canada (\href{http://archive.cra.org/reports/forsythe.html}{archive.cra.org/reports/forsythe.html}). 

For each department, the dataset provides a complete list of regular faculty for the 2011--2012 academic year, and for each of the 5032 faculty in this collection, it provides partial or complete information on their education and academic appointments, obtained from public online sources, mainly r\'esum\'es and homepages. Of these, we selected the 2583 faculty who both received their Ph.D.\ from and held their first assistant professorship at one of these institutions, and for whom the year of that hire is known and occurred in 1970--2011. The first requirement ensured that we modeled the relatively closed North American faculty market; roughly 87\% of computing faculty received their Ph.D.\ from one of the Forsythe institutions, and past analysis has shown that Canada and the United States are not distinct job markets in computer science~\cite{clauset2015systematic}. A number of faculty were removed in this step because the location of their first assistant professorship was not known; these were mainly senior faculty. 

The first dataset also provides a ranking of institutional prestige $\pi$, derived from patterns in the PhD-to-faculty hiring network between departments. In short, $\pi$ is a consensus of ordinal rankings (lower is better) in which prestige is defined recursively: prestigious departments are those whose graduates are hired as faculty in prestigious departments. Networks, code, and rankings are available in Ref. \cite{clauset2015systematic}.

The second dataset, constructed around the first one, is a complete publication history as listed in the Digital Bibliography and Library Project (DBLP; \href{http://dblp.uni-trier.de}{dblp.uni-trier.de}), an online database that provides open bibliographic information for most journals and conference proceedings relevant to computing research, using manual name disambiguation as necessary. For each paper in a faculty's publication history, we recorded the paper's title, author list (preserving author order), and year of publication. By following this procedure, we collected data for 200,476 publications which covered 2453 (95.0\%) faculty in our sample. Of those, we manually collected records of all peer-reviewed conference and journal publication histories from the publicly available curricula vitae (CVs) of 109 faculty, a randomly selected 10\% of the 1091 faculty with career lengths between 10 and 25 years, providing a benchmark dataset to evaluate the accuracy of DBLP data (Supplemental Text~\ref{supp:cv_data}).

Our combined dataset consists of the career trajectories of these 2453 tenure-track faculty as of 2011--2012: each professor's publicly-accessible metadata, their time-stamped Ph.D. and employment history, and the annotated time series of their publications. We note that this dataset does not include information on faculty who have retired or left academia prior to 2011. Implications of these data limitations for the conclusions that can be drawn from our analyses are explored in the Discussion section. Finally, this study was not reviewed by an institutional review board because all data used were collected from publicly available sources. All results are presented anonymously or in aggregate to avoid revealing personally identifiable information about individual scientists.


\section*{Results}

\subsection*{General Trends in Productivity}

Two broad trends characterize scholarly productivity in academic computer science. First, publication rates have been increasing over the past 45 years, and second, higher publication rates are correlated with higher prestige. These two observations are intertwined and underpin a number of subsequent analyses, so we explore them briefly in more depth. 

\begin{figure}[h!]
	\centering
	\includegraphics[width=0.95\linewidth]{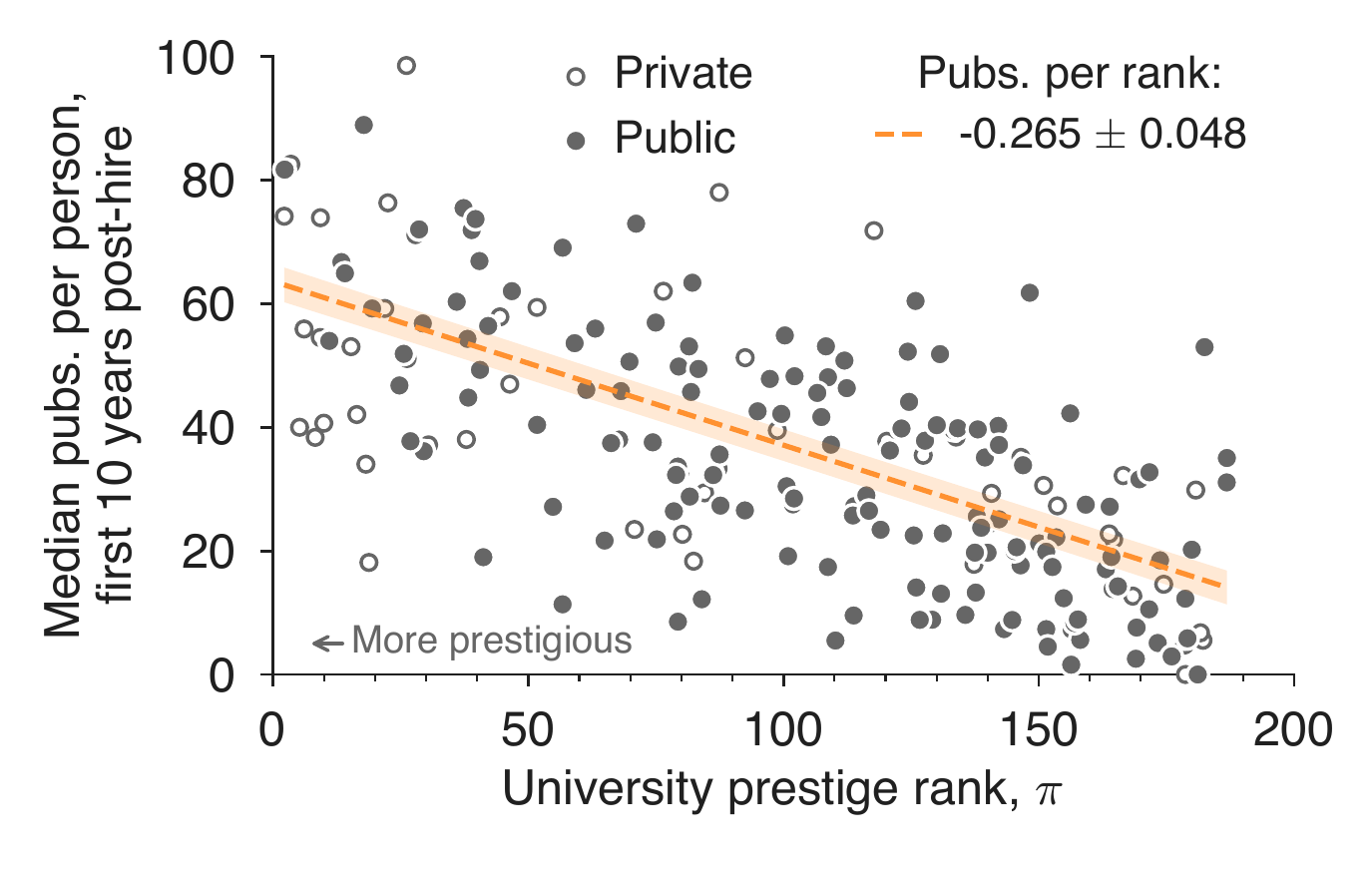}
	\caption{{\bf Publications correlate with institution prestige.} {\normalfont Dots indicate median number of publications per person per institution for researchers' first ten years post-hire, adjusted for growth in publication rates over time (see Supplemental Figs.~\ref{fig:dblp_data}--\ref{fig:growth}
	and Supplemental Text~\ref{supp:data}) 
	and ordered by institutional prestige, $\pi$ \cite{clauset2015systematic}. Effects of prestige are similar for private (open circles) and public (closed circles) institutions ($p=0.146$, $t$-test; see text), increasing at a rate of nearly 2.7 publications per 10-rank improvement in prestige. Shaded region denotes the 95\% confidence interval for least squares regression.}}
	\label{fig:pubsvrank}
\end{figure}

Past studies have found that researchers at more prestigious institutions tend to be more productive \cite{crane1965scientists,zuckerman1970stratification,pelz76organizations,long1978productivity,reskin1979academic,allison1990departmental}. Our data corroborate this finding, but we also find that the typical productivity advantage associated with greater prestige holds regardless of whether an institution is public or private, for both early-career publications (first ten years; Fig.~\ref{fig:pubsvrank}) and lifetime publications\ (Supplemental Fig.~\ref{fig:supp_pub_priv}).
Regressing the median number of time-adjusted publications (see below) among faculty in a department against departmental prestige indicates that the relationships between prestige and productivity are statistically indistinguishable for public and private institutions, with expected increases in the first decade of a career of roughly 2.7 publications for every 10-rank improvement in prestige. 
In fact, when comparing public and private institutions, neither the prestige-productivity slope nor productivity overall is significantly different ($p\!=\!0.150$, $0.148$, respectively, two-tailed $t$-test), contradicting the conventional wisdom that private universities enjoy a productivity advantage over public ones. The conventional wisdom is likely skewed by a focus on elite departments, as 8 of the top 10 computer science departments are private \cite{clauset2015systematic}, but in fact, private institutions are distributed evenly across all ranks. Expanding this analysis to include lifetime publications increases the prestige-publication slope to 3.28 publications per 10-rank improvement in prestige but does not alter the non-significance of public/private status ($p\!=\!0.714$, $0.346$, two-tailed $t$-test).

Past studies have also found that publication rates have increased over time \cite{dey1997changing,larsen2010rate}. 
However, prior to investigating whether changes in publication rates apply to computer science, we used the manually collected CV data to probe the extent of DBLP's coverage. Indeed, the fraction of publications indexed by DBLP is non-uniform over time, increasing linearly from around 55\% in the 1980s to over 85\% by 2011 ($R^2$=0.685 $p\!<\!0.001$, two-tailed $t$-test; see Supplemental Figure~\ref{fig:dblp_data} and Supplemental Text~\ref{supp:data}). Because DBLP's coverage of the published literature varies over time, in the analyses that follow we use data from hand-collected faculty CVs whenever possible, and otherwise apply a statistical correction to DBLP's data in order to account for its lower coverage.

Knowing already that there are substantial differences in productivity by prestige, we separated universities by prestige into five groups of approximately equal size, and investigated whether the growth of publication rate varies by prestige. We find that the average number of publications per person produced in each calendar year has been increasing at all five strata of prestige at rates between 0.72 and 1.23 publications per decade, for 45 years (Supplemental Figure~\ref{fig:growth}).
Because we have used data from hand-collected faculty CVs to adjust DBLP-derived paper counts for DBLP's steadily improving coverage over time, these estimated growth rates represent a real increase in publication rates over this 40 year period. Moreover, the observed steady increase in productivity is not uniform across prestige, and the difference between production growth rates between higher and lower prestige departments have widened slightly but significantly over time ($p\!<\!0.05$, two-tailed $t$-test). In other words, prestigious and non-prestigious institutions have contributed to the overall growth at different rates. Not only are there small but significant differences in productivity by prestige (Figure~\ref{fig:pubsvrank}) but those differences are slowly growing (Supplemental Figure~\ref{fig:growth}).

To investigate the productivity patterns of individual researchers and test the conventional narrative of rapidly rising productivity followed by a gradual decline, for the remainder of this paper we focus on time series of individual productivity. However, due to both the observed growth in productivity, and the variability in DBLP coverage, it would be misleading to directly compare a 1975 publication with a 2011 publication. Thus, hereafter we use ``adjusted'' publication counts, which corrects the raw DBLP counts to account for both the changing DBLP coverage and the increasing mean publication rate over time (Supplemental Text~\ref{supp:data}).
All publication counts are hence reported as 2011-equivalent counts.

\subsection*{Individual Productivity Trajectories}

Examining the productivity trajectories of individual researchers, we find that they too exhibit substantial and significant differences in their publication rates. Early studies of scholarly productivity noted profound imbalance in the number of articles published by individual researchers \cite{lotka1926frequency,shockley1957statistics}.
Cole \cite{cole1979fair} and Reskin \cite{reskin1978scientific} in the 1970s noted that about 50\% of all scholarly articles were produced by about 15\% of the scientific workforce. Our data reflect similar levels of imbalance, with approximately half of all contributions in the dataset authored by only 20\% of all faculty. Stratifying by decade, however, the Gini coefficients for productivity imbalance have been declining, from 0.62 in the 1970s to 0.40 in the 2000s (See Supplemental Figure~\ref{fig:imbalance}). 
This trend persists when researchers are restricted to only the publications within the first five years of their careers. 

There are several possible explanations for the trend of decreasing inequality in individual productivity. For instance, the lower end of the productivity distribution could have become relatively more productive over time, perhaps as more institutions shifted focus from teaching to research. Or, it may reflect a strengthening selective filter on highly productive faculty, perhaps as community expectations for continual productivity rose. It may also reflect non-uniform errors in the DBLP data, although the correction for DBLP coverage should account for these (Supplemental Text~\ref{supp:data}).

\begin{figure}[t]
	\centering
	\includegraphics[width=1\linewidth]{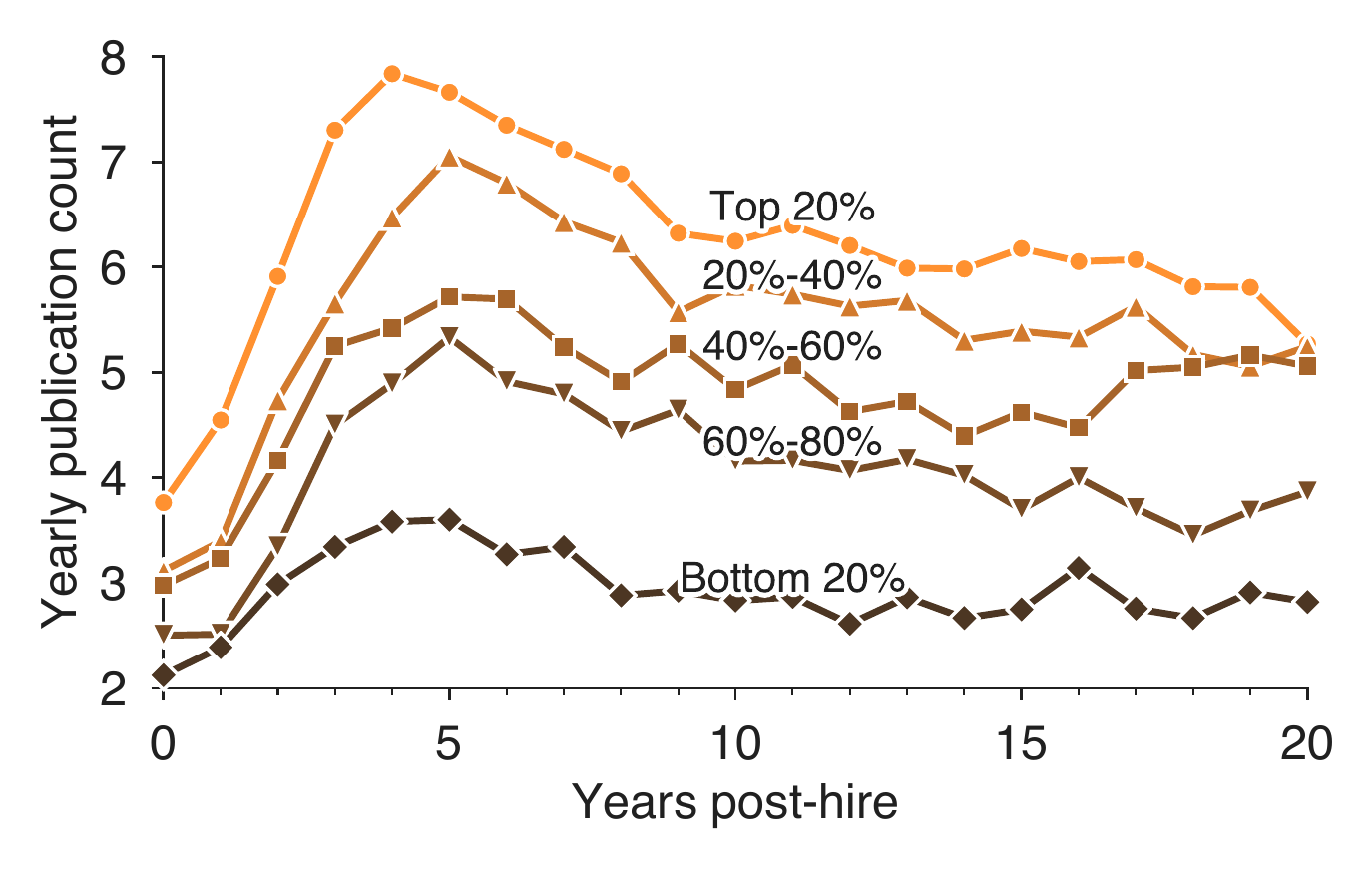}
	\caption{{\bf Average publications follow conventional narrative across prestige.} {\normalfont 	
	Five average productivity curves are shown, partitioning universities according to prestige rank $\pi$ such that each quintile represents approximately 20\% of all faculty in the full dataset. Averages over researchers at all levels of institutional prestige follow similar productivity trajectories, in agreement with the conventional narrative, but at differing scales of output.}}
	\label{fig:prod_curves_by_rank}
\end{figure}

We now focus on testing the conventional productivity narrative that has been described in various disciplines and at many points in time \cite{lehman1954men,dennis1956age,cole1979age,diamond1986life,horner1986relation,simonton1997creative,levin1991research,bayer1977career,symonds2006gender,quetelet1835homme,kaplan2000theory}: productivity climbs to a peak and then gradually declines over the course of the researcher's career. Across computer science faculty, we find that the average number of publications per year over a faculty career is highly stereotyped (Figure~\ref{fig:prod_curves_by_rank}), with a rising productivity that peaks after around 5 years, declines slowly for another 5 years, and then remains roughly constant for any remaining years. Although departmental prestige correlates with productivity in several ways (Figures~\ref{fig:pubsvrank} and \ref{fig:growth}),
it does not alter this stereotypical pattern, which appears essentially unchanged across departments with different levels of prestige, except for a roughly constant shift up as prestige increases (Figure~\ref{fig:prod_curves_by_rank}).

The suggestion that productivity grows in the early years of a career has intuitive appeal. Professors settle into their research environments, begin training graduate students, and build their cases for promotion and tenure. Similarly, many reasons have been suggested for why productivity might decrease after promotion, including increased service and non-research commitments, declining cognitive abilities, and increased levels of distraction from outside work due to health issues and childcare obligations \cite{fox1983publication}.
Although an average over faculty appears to confirm the stereotyped trajectory of rapid growth, peak, and slow decline, it does not reveal whether this average is representative of the many individual trajectories it averages over, nor does it show how much diversity there might be around the average, and whether that diversity correlates with other factors of interest.

\begin{figure}[t]
	\centering
	\includegraphics[width=0.95\linewidth]{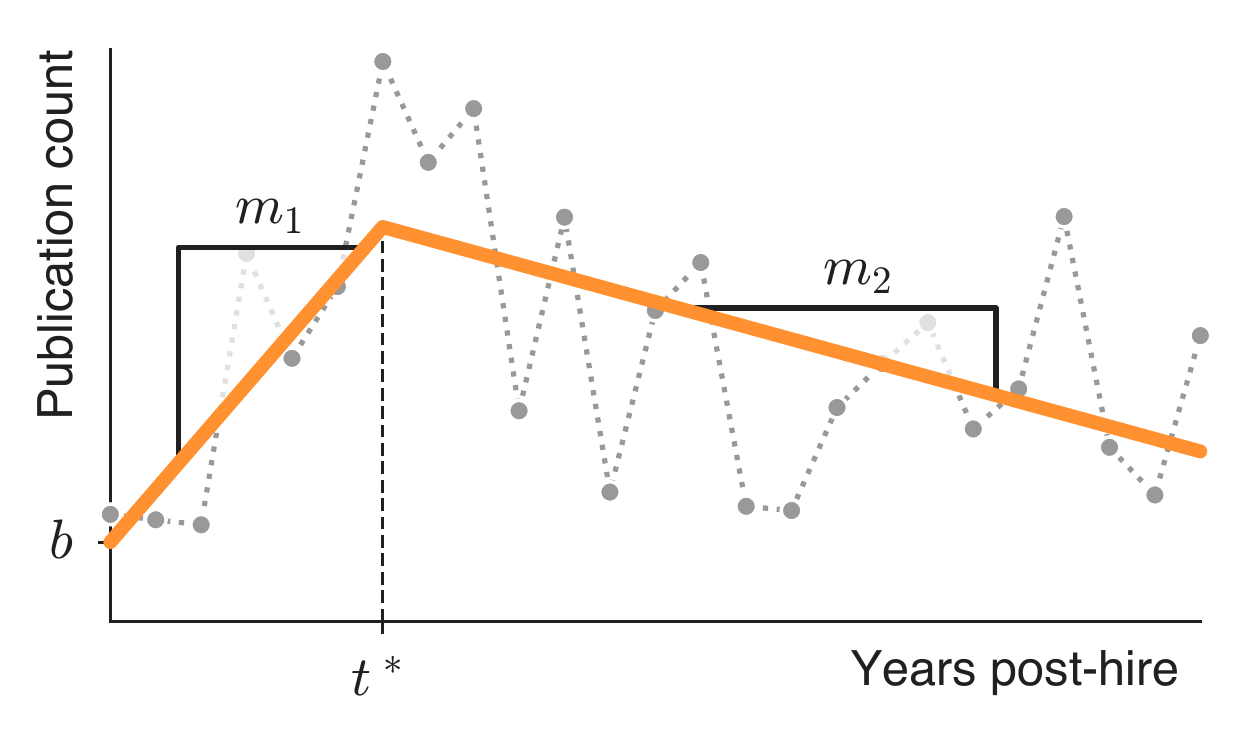}
	\caption{{\bf Example trajectory and piecewise model.} {\normalfont Dots represent empirical annual publications. Orange line shows best fit of piecewise linear model [Eq.~\eqref{eqn:piecewise}] with slopes $m_1$, $m_2$, change point $t^*$, and intercept $b$ annotated.}}
	\label{fig:piecewise}
\end{figure}

To characterize the productivity pattern within an individual career, we fit a simple stereotypical model of productivity over time to the number of papers published per year,
\begin{equation}
	\label{eqn:piecewise}
f(t) = 
\begin{cases} 
      b+m_1 t & 0\leq t\leq t^* \\
      b+m_1 t^* + m_2 (t-t^*) & t>t^*
\end{cases}
\end{equation}
a piecewise linear function in which $t^*$ is the change point between the two lines, $m_1$ and $m_2$ are the rates of change in productivity before and after the change point, respectively, and $b$ is the initial productivity (Figure~\ref{fig:piecewise}). We apply this model to the $N\!=\!1091$ faculty who have been employed for 10--25 years. By fitting these four parameters to each individual's publication trajectory, we map that trajectory into a low-dimensional description of its overall pattern (fitting done by least squares; see Supplemental Text~\ref{supp:leastsquares} for optimal numerical methods and \ref{supp:modeling} for detailed discussion of statistical models). 

However, prior to interpreting the distributions of parameters, we subjected each trajectory to two additional tests to ensure that its best-fit parameters were meaningful. First, to avoid overfitting linear trajectories with a piecewise linear model, we performed model selection, asking whether AIC with finite-size correction favored a straight line or the more complex $f(t)$ (see Supplemental Text~\ref{supp:model_selection}). 
This process conservatively selected only 33.3\% ($N\!=\!363$) of researchers who are more confidently modeled by the piecewise function.

Second, to address the possibility that a researcher's best-fit parameters may be sensitive to small changes in the years of their publications, we conducted a sensitivity analysis in which we repeatedly re-fit model parameters to productivity trajectories, adding a small amount of noise to shift some publications into adjacent years (see Supplemental Text~\ref{supp:sensitivity}).
This procedure places each professor's noise-free trajectory within a distribution of nearby noisy trajectories, enabling two different (but ultimately concordant) analyses. The primary sensitivity analysis focuses on individual faculty, computing whether the parameters of each professor's noise-free trajectory are similar to their noisy distribution. This approach revealed that a majority (77.2\%) of trajectories are well represented by their noise-free parameters, each consistently falling into the same region of parameter space for over 75\% of resampled trajectories. We refer to these trajectories as ``stable'' in subsequent analyses, meaning that their noise-free parameters are representative and interpretable. The alternative sensitivity analysis focuses on the population of faculty, combining all noise-free trajectories with their noise-added distributions into a single expanded ensemble of conceivable productivity trajectories (see Supplemental Text~\ref{supp:sensitivity}). 
Although this ensemble is unable to support analyses of individual faculty, we use it to corroborate the findings that follow. Combining the individual stability and AIC criteria, we find that 32.3\% ($N\!=\!352$) of researchers possess productivity trajectories that are both stable and non-linear. All analyses and discussions of model parameters hereafter refer to stable, non-linear trajectories unless otherwise noted.

\begin{figure*}[t!]
	\centering
	\includegraphics[width=1.0\linewidth]{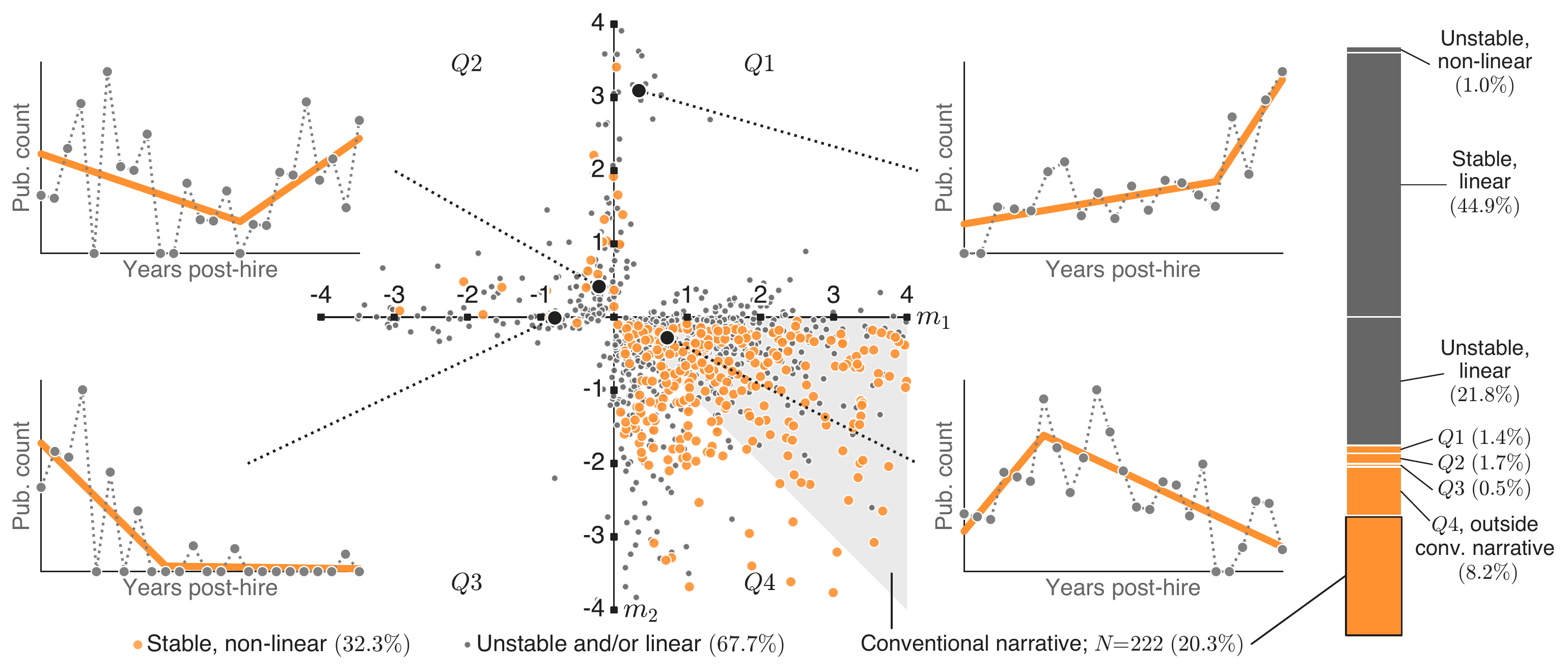}
	\caption{{\bf Distribution of individuals' productivity trajectory parameters.}  {\normalfont
	Diverse trends in individual productivity fall into four quadrants based on their slopes $m_1$ and $m_2$ in the piecewise linear model Eq.~\eqref{eqn:piecewise}. Subfigures show example publication trajectories to illustrate general characteristics of each quadrant. The shaded triangular region corresponds to the conventional narrative of early increase followed by gradual decline. Color distinguishes trajectories in two classes: those that are stable and non-linear (orange), and those that are either unstable or linear (gray). The plot on the right describes how researchers are distributed within these two classes.}}
	\label{fig:piecewise_shapes}
\end{figure*}

The narrative of ``early growth in productivity, followed by a slow decline'' implies four conditions on the inferred parameters: while the conditions of growth ($m_1>0$) and decline ($m_2<0$) are straightforward, we interpret ``early growth'' to mean that inferred peak productivity comes within the first decade after hiring ($t^* \leq 10$) and ``slow'' to mean that the slope of decline is smaller in magnitude that the slope of growth $(|m_2| < m_1)$. After fitting individual trajectory models to the 1091 faculty in our sample, we find that only 20.1\% follow the stereotypical trajectory. Even dropping the aforementioned restriction on $t^*$ increases the fraction meeting the stereotype to only 20.3\%. To ensure that these results were not sensitive to our definition of stability in the presence of noise, we generated an ensemble with 200 noise-added trajectories for each professor (see Supplemental Fig.~\ref{fig:noisy_fits} and Supplemental Text~\ref{supp:sensitivity}), subjected each to the AIC criterion for nonlinearity, and found that only 19.7\% of ensemble trajectories are reliably categorized as adhering to the conventional narrative. In other words, the average trajectory, which has been held up as established fact for more than 50 years, describes the behavior of only a minority of researchers, while a large majority of researchers follow qualitatively different trajectories. 

Publication trajectories can be divided into four general classes based on the signs of the two slope parameters, $m_1$ and $m_2$, corresponding to the quadrants shown in Figure~\ref{fig:piecewise_shapes}. Individual trajectory shapes exhibit substantial diversity, spanning all four quadrants. Even among faculty whose publication rates grew and then declined (lower right quadrant, 28.6\%), the conventional narrative only includes the 20.3\% of individuals whose rate of growth exceeds their rate of decline ($m_1>|m_2|$; shaded region,  Figure~\ref{fig:piecewise_shapes}). Additionally, researchers were distributed similarly across the four quadrants, comparing parameters extracted from DBLP data versus hand-collected CV data ($p=0.14$, $\chi^{2}$), confirming that the dispersion shown in Figure~\ref{fig:piecewise_shapes} represents the true diversity of careers.

The cloud of faculty trajectory parameters shown in Figure~\ref{fig:piecewise_shapes} does not naturally separate into coherent clusters. In their absence, what are the covariates that predict which region of the plot an individual is likely to occupy? First, early-career growth rate of yearly publications $m_1$ is significantly correlated ($p\!<\!0.001$, $t$-test) with the prestige of researchers' employing institutions. This is particularly true for researchers at ``elite" institutions, which we define as being in top 20\% of universities according to prestige rank and adjusting for number of faculty (same partitions as Figure~\ref{fig:prod_curves_by_rank}). Specifically, researchers productivity grows by a median of 2.02 additional papers per year at elite institutions compared with 1.19 for others  ($p\!<\!0.001$, one-tailed Mann-Whitney). Perhaps as a result---what goes up must come down---the slope after the point of change, $m_2$, correlates significantly with prestige and is more negative for researchers at higher-ranked institutions, compared to those at lower-ranked institutions ($p\!<\!0.05$, $t$-test). Additionally, researchers who received their doctorates from elite institutions exhibit faster early-career growth than those who trained at lower-ranked institutions ($p\!<\!0.05$, one-tailed Mann-Whitney).

Second, the early-career initial productivity $b$ is significantly higher for faculty who graduated from elite departments ($p\!<\!0.005$, one-tailed Mann-Whitney). We also find that researchers who place into elite departments or who have postdoctoral experience tend to start out more productive, however these differences are not statistically significant ($p\!>\!0.05$, Mann-Whitney).  These findings regarding $m_1$ and $b$ combine to suggest that current academic environment correlates with---and perhaps influences---productivity, while prior academic environment does not. Finally, faculty at top-ranked departments are statistically no more or less likely to be found within this triangular region, a result robust to alternatives cutoffs for ``top ranked'' institutions. 

The relationship between trajectories and gender is more complicated. First, trajectories of male and female researchers were similarly distributed across the four quadrants ($p\!=\!0.94$, $\chi^{2}$), and gender was uncorrelated with the likelihood of meeting the four criteria of the canonical narrative ($p\!=\!0.39$, $\chi^{2}$). Further, within this canonical subset, the women's initial productivity grew at a rate indistinguishable from the men's ($p\!=\!0.15$, Mann-Whitney) and peaked in similar years ($p\!=\!0.305$, Kolmogorov-Smirnov). Women's initial productivity, however, was 0.46 publications lower than the men's ($p\!=\!0.032$, Mann-Whitney) in general, and this difference exists in spite of the fact that men and women in this subset trained and were hired at similarly-ranked institutions ($p\!>\!0.05$, Kolmogorov-Smirnov), and completed postdoctoral training at similar rates ($p\!=\!0.89$, $\chi^{2}$).

\begin{figure}[t]
	\centering
	\includegraphics[width=0.825\linewidth]{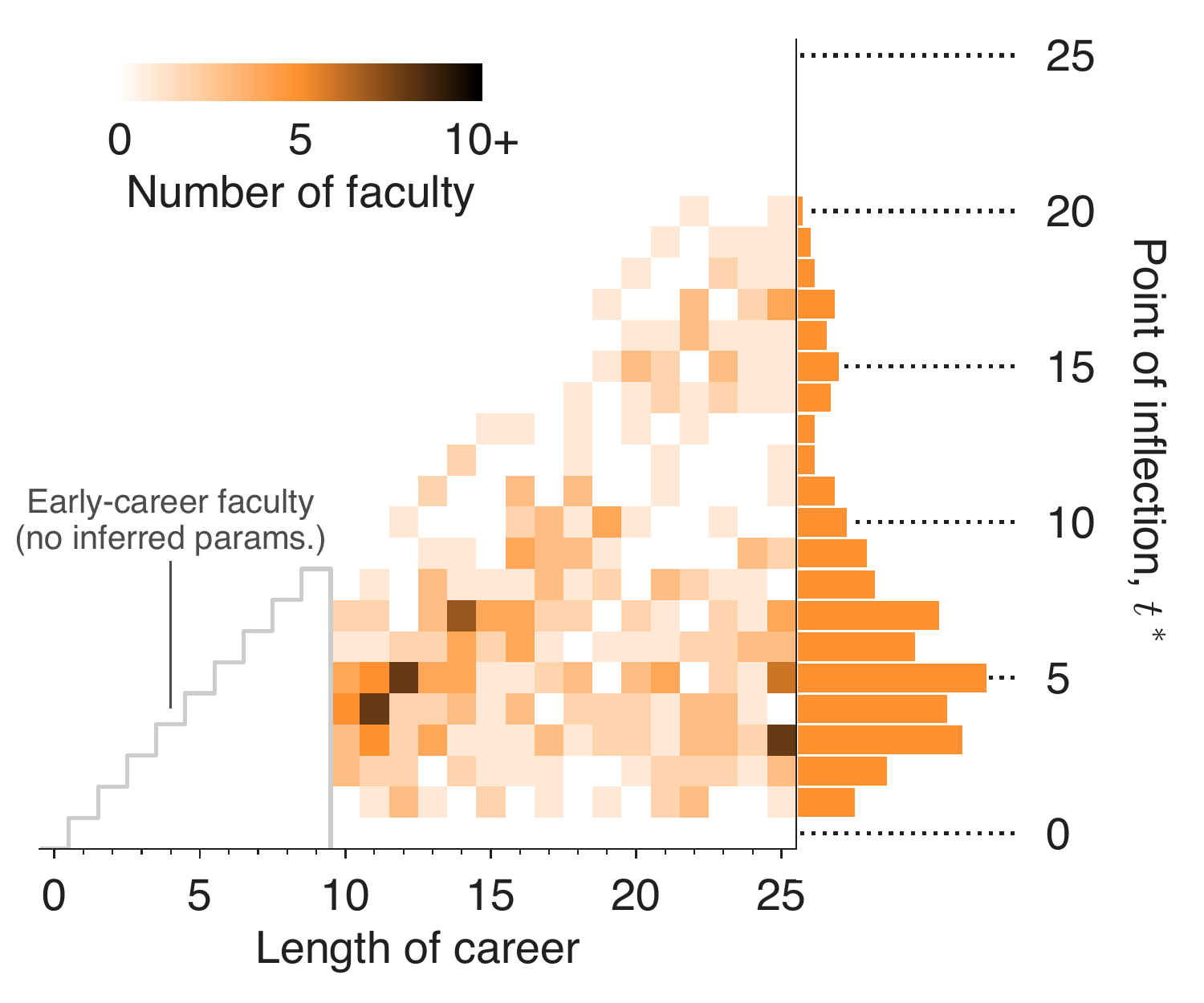}
	\caption{{\bf Heat map of researchers' inferred change points.} {\normalfont Each researcher's inferred change-point parameter $t^*$ is plotted as a heat map, sorted by the length of their career in our dataset and restricted to individuals whose productivity trajectories are both stable under the addition of noise (see text) and better modeled by Eq.~\eqref{eqn:piecewise} than a straight line, determined by AIC (see Supplemental Text~\ref{supp:model_selection})}.} 
	\label{fig:change}
\end{figure}

The change point within a career may indicate regime shifts in productivity, regardless of which type of trajectory an individual may follow. While the change-point parameter $t^*$ does not correlate with the other parameters of $f(t)$, its distribution reveals that for most faculty, the inferred change point in productivity rates occurs at approximately year 5. Figure~\ref{fig:change} translates each selected faculty member's career length and inferred change point into an ordered pair, creating a heat map of career change points. Shown in the accompanying marginal distribution, the modal value for $t^*$ is year five with the median at 6 years, closely preceding tenure decisions at most institutions. Nevertheless, there is still rich diversity in career transitions, and the average remains misleading as the descriptor of a majority of individuals. In particular, faculty at the top 20\% of institutions have significantly earlier $t^*$ than the remaining 80\%, with medians of 4.1 years and 6.4 years, respectively ($p\!<\!0.001$, Mann-Whitney). There is no such difference between the faculty whose doctorates are from the top 20\% of institutions and those whose doctorates are from the remaining 80\% (medians of 5.9 versus 6.0 years; $p\!=\!0.37$).

\begin{figure}[t]
	\centering
	\includegraphics[width=0.825\linewidth]{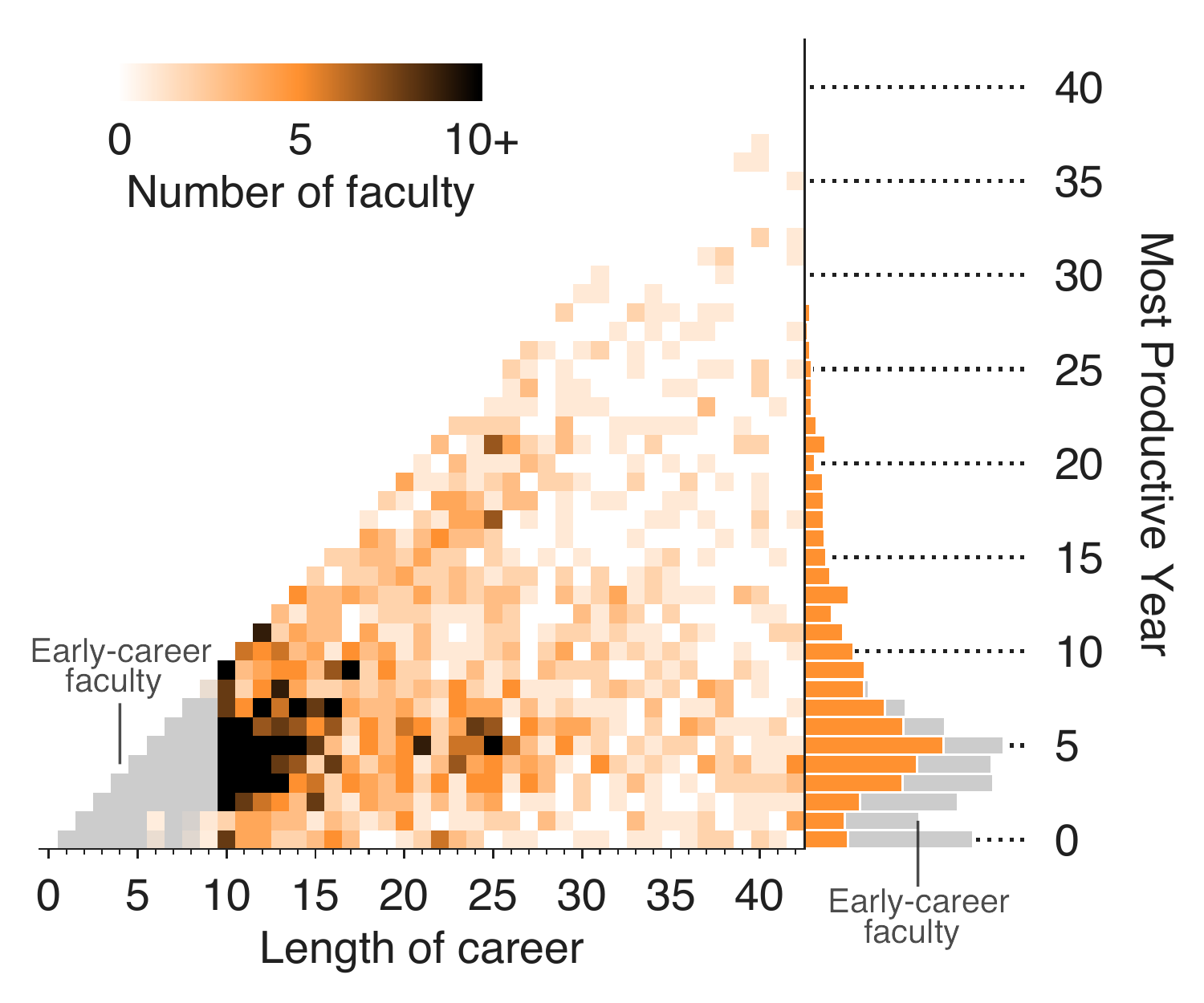}
	\caption{{\bf Heat map of researchers' most productive years.} {\normalfont Each researcher's most productive year (empirically; not model fit) is plotted as a heat map, sorted by the length of their career in our dataset. White box indicates researchers with fewer than 10 years of experience, whose most productive year is necessarily early. The marginal distribution (right) shows the empirically most productive year for all faculty in the dataset, separated by early career (first 10 years; gray) or later career (orange). The most common peak-productivity year is year 5, and only about half of senior faculty exhibit peak productivity in year 5 or earlier.}}
	\label{fig:most_prod}
\end{figure}

The trends and diversity observed in $t^*$ distributions remain true even when models are avoided entirely. A direct empirical examination of all DBLP and CV publication time series reveals that a computer science professor's productivity is also most likely to peak in the fifth year, yet peak productivity can nevertheless occur in any year of a professor's career (Figure~\ref{fig:most_prod}). While the marginal distribution shows that 41.9\% of faculty have their peak productivity within the first 6 years, with the modal peak year in year 5, there is substantial variance. Note, for example, that individuals along the bottom of Figure~\ref{fig:most_prod} published the most in their first year as faculty, while individuals along the diagonal published the most in their most recent recorded year as faculty. 

\subsection*{Transitions in Authorship Roles}

\begin{figure}[t]
	\centering
	\includegraphics[width=0.8725\linewidth]{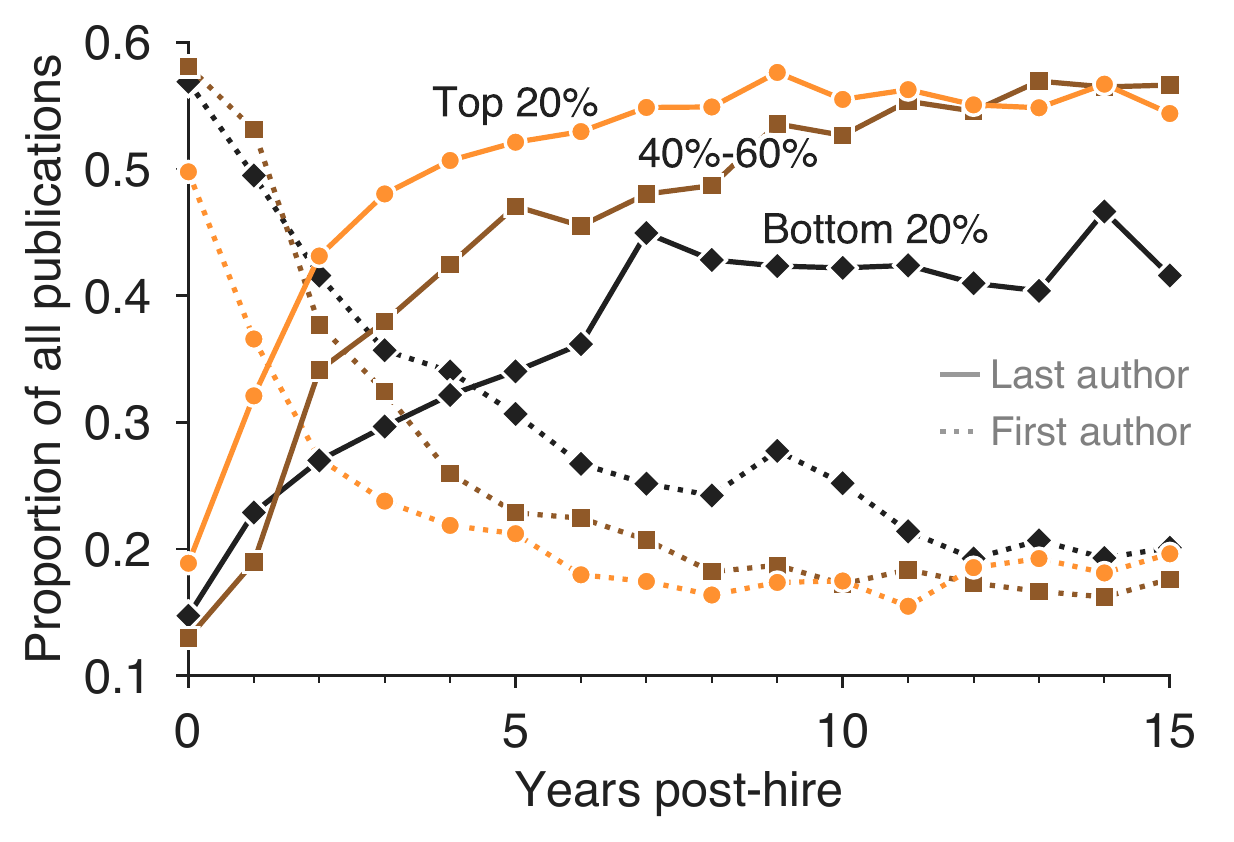}
	\caption{{\bf Early-career transitions in authorship roles.} {\normalfont The average proportion of first-author (dotted lines) and last-author papers (solid lines) as fraction of total, as a function of career age, separating researchers at institutions in the top, middle, and bottom quintiles according to prestige rank. Single-author publications are counted as first-author publications. On average, researchers at more prestigious institutions transition more quickly into senior-authorship roles.}}
	\label{fig:lofaps}
\end{figure}

Finally, other transitions exist that are not quantifiable in publication counts alone, yet these are surprisingly well synchronized with the transitions noted above. As faculty train graduate students, their roles ordinarily shift from lead researcher to senior advisor or principal investigator, and this transition is commonly reflected in a shift from first author to last author. While common, this first/last convention is not universal. For example, papers in theoretical computer science typically order authors alphabetically, so the relative position of these researchers in the author list will not exhibit any consistent pattern over a career. To investigate career-stage transitions in author position, we first identified the set of journals or conferences that list authors alphabetically by computing whether each venue's authors are alphabetized significantly more often than is expected by chance ($\alpha\!=\!0.05$) and exceeding twice the expected rate (see Supplemental Text~\ref{supp:alphabetized}). 
These conditions selected 11.2\% of publication venues, accounting for 15.4\% of all papers in the dataset, which we manually verified includes all top theoretical computer science conferences and excludes all top machine learning and data mining conferences. We then discarded these alphabetically biased venues from the following analysis. The remaining data show clear evidence of a progressive shift toward last-authorship position over time, with the relative first/last proportion reaching stability around year eight (Figure~\ref{fig:lofaps}). Interestingly, the onset of this change is earlier among faculty at high-prestige institutions, and their average proportion of last-author papers is significantly higher than those of other faculty, consistent with a hypothesis that faculty at elite institutions tend to begin working with students earlier and have larger or more productive research groups.

\begin{figure}[t]
	\centering
	\includegraphics[width=0.68\linewidth]{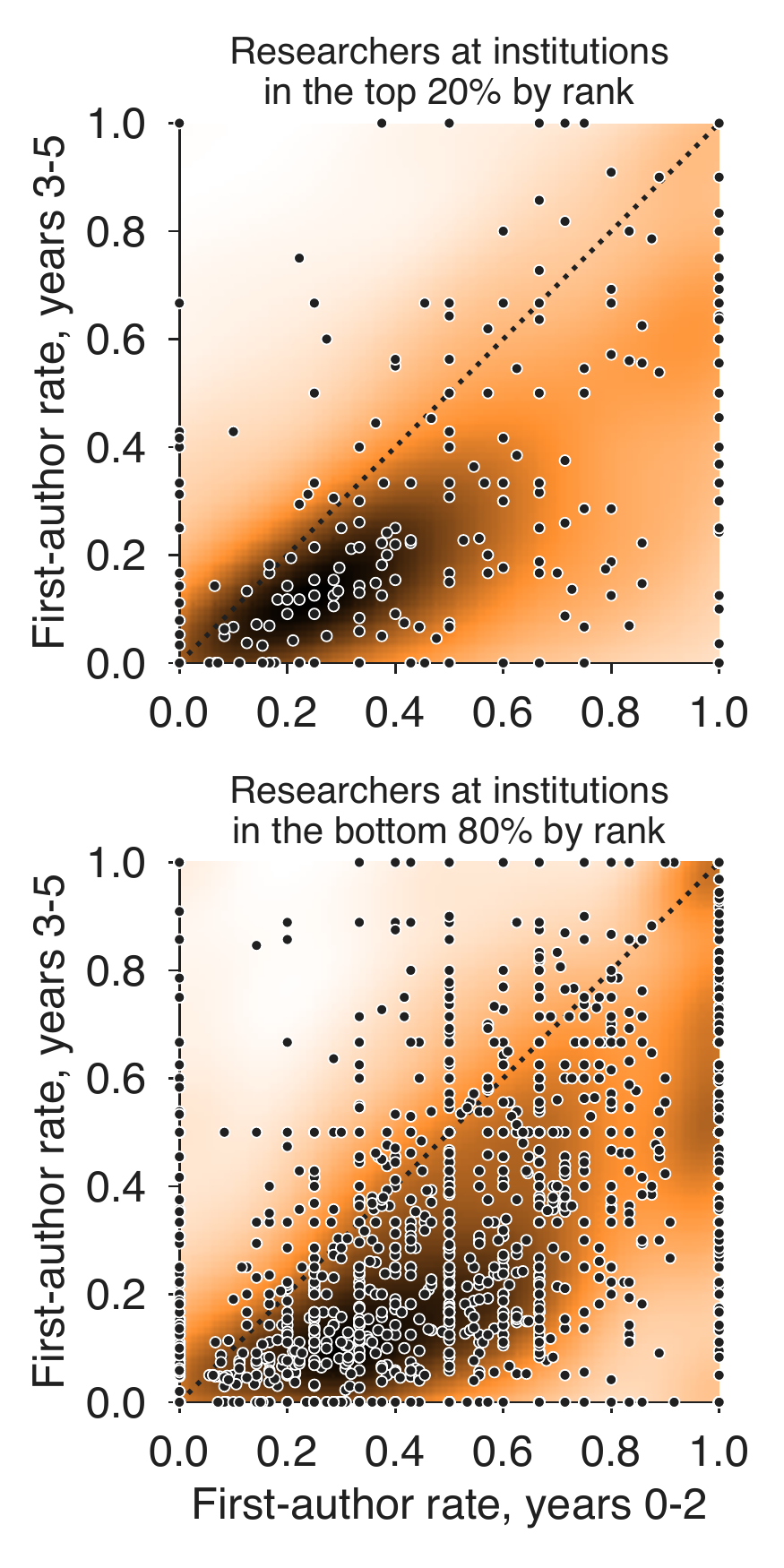}
	\caption{{\bf First-author publication rates.} {\normalfont First-author publications as fraction of total in the first three years post-hire, and the three years thereafter, shown separately for researchers who placed at an institution in the top 20\% by rank (top) and researchers placing outside of the top 20\% (bottom). Individual researcher data are plotted as points on top of a corresponding heatmap in which darker color denotes higher density by Gaussian kernel density estimation. Researchers at all levels of prestige tend to move out of first-authorship roles during this period, though researchers at more prestigious institutions transition more completely by years 3--5 than others.}}
	\label{fig:faps}
\end{figure}

As with the aggregate trend in productivity over a faculty career (Figure~\ref{fig:prod_curves_by_rank}), the transition from first- to last-author publications (Figure~\ref{fig:lofaps}) is based on averaging across many faculty and thus may not reflect the pattern of any particular individual. To characterize individual performances, we compared the fraction of first-authored papers in the first three years post-hire to the same fraction in the second three years, for faculty with careers longer than six years ($N\!=\!2036$). A substantial drop in this fraction across these two periods would be consistent with the average trend reflecting individual patterns. For this analysis, we treated single-author papers as first-author publications. Overall 70.1\% of researchers undergo this transition, publishing a larger fraction of first-author publications in the first three years of their faculty career than in the second three years. These fractions are consistent for faculty at top-50 institutions (70.2\%) and those at other institutions (70.1\%), but individuals at top-ranked institutions appear to make the transition more quickly and completely by the end of the six-year period (Figure ~\ref{fig:faps}). In spite of these trends, there remains substantial diversity among first/last author transitions, reinforcing the notion that averages may be poor descriptors of many individuals. 

\section*{Discussion}

The conventional narrative of faculty productivity over a career is pervasive, with repeated findings reinforcing a canonical trajectory where productivity rises rapidly to a peak early in one's career and then declines slowly \cite{lehman1954men,dennis1956age,cole1979age,diamond1986life,horner1986relation,simonton1997creative,levin1991research,bayer1977career,symonds2006gender,quetelet1835homme,kaplan2000theory}. This narrative shapes expectations of faculty across career stages, and publication counts have been shown to impact both tenure decisions \cite{cole1967scientific,massy95}, and the ability to secure funding for future research \cite{stephan1996economics}. In this study, we showed that the conventional narrative, while intuitive, and certainly applicable to averages of many professors, is a remarkably inaccurate description of most professors' trajectories. By applying a simple piecewise-linear model to a comprehensive dataset of academic appointment histories and publication records, we found that only about one fifth of tenured or tenure-track computer science faculty resemble the average, regardless of their department's prestige. 

While diverse, some aspects of a trajectory are nevertheless partially predictable. For example, although the diversity of trajectories remains unaffected, productivity does tend to scale with prestige: researchers who graduated from or were hired by top-ranked institutions are significantly more productive at the onset of their careers, and, furthermore, productivity of high-prestige faculty tends to grow at faster rates and achieve higher peaks than researchers employed by other institutions. Together, these results support previously suggested hypotheses that top-ranked universities both select for and facilitate productivity \cite{allison1990departmental}. In fact, our results suggest that the early-career transition to leadership roles, a phenomenon also found other disciplines \cite{duch2012possible}, takes place more quickly at top-ranked institutions, further implicating facilitation effects in addition to selection.

The relationship between productivity trajectories and gender is complicated and requires careful study. Gender has been shown to correlate with differences in productivity across fields \cite{long1992measures,xie1998sex,fox2005gender}, but these relationships are complicated by prestige \cite{clauset2015systematic} and have also changed over time \cite{way2016gender}. Other work has uncovered differences in collaboration patterns between subfields \cite{zeng2016differences}, as well as productivity differences that depend on both student and advisor genders \cite{pezzoni2016gender}. Here, we found that men and women follow the canonical productivity narrative at equal rates. However, among those who do, we found significant differences in initial and peak productivities between men and women. Given the complications revealed in past studies, the extent to which these differences reflect inequalities, past or present, and contribute to women?s underrepresentation in computer science is an important topic of research and warrants future exploration.

Within the space of career trajectories, there is a noticeable tendency toward peak productivities and shifts in publication rates around 5 years after beginning as faculty. This is surely not a coincidence, given the fundamental role of tenure as a change point within the typical academic career, after which the total number of hours worked does not substantially change, but the time devoted to service tends to dramatically increase, with concomitant decreases in research and grant-writing \cite{link2008time}. However, our data cannot yet say how, from a mechanistic perspective, the existence of tenure requirements drives faculty to change or shape their productivity before or after promotion. If anything, the results in this paper make clear that there are numerous ways in which computer scientists meet promotion requirements, not all of which necessarily involve publishing a large number of papers. Indeed, in parallel with career shapes more broadly, there remains broad diversity in the distributions of productivity peaks and change points. This diversity in overall production, combined with the observation that an individual's highest impact work is equally likely to be any of his or her publications \cite{sinatra2016quantifying}, implies there are fundamental limits to predicting scientific careers \cite{clauset2017data}.

Computer science is, itself, a multifaceted field, and previous studies of the DBLP dataset revealed that productivity rates differ by subfield \cite{way2016gender}. This observation, coupled with the menagerie of fluctuating trajectories revealed here, may suggest that year-to-year differences in individual trajectories are related to which subfields a researcher studies. Past work has revealed a first-mover advantage associated with entry into a rapidly growing field \cite{newman2009first}, so changes to individual research interests may contribute to noisy trajectories, particularly if they coincide with concentrated growth of popular new subfields. 

Larger and higher-resolution data sets may improve our ability to identify expanding new subfields and other factors that could explain or predict trajectories. Although DBLP has the advantage of covering computer science journals and peer-reviewed conferences alike, we found that its coverage of those venues was incomplete in predictable ways. By manually collecting CV data for 10\% of the scattered trajectories shown in Figure~\ref{fig:piecewise_shapes}, we adjusted DBLP data for missing publications and established the rate at which publishing rates have grown since 1970. Trajectories derived from DBLP data and benchmark CV data were statistically indistinguishable from each other. Investigations of productivity trajectories outside computer science will lack the field-specific DBLP database and may require additional calibration, name disambiguation, and data deduplication.

The misleading narrative of the canonical productivity trajectory is not likely to be unique to computer science. The rich diversity revealed here demands a reevaluation of the conventional narrative of careers across academia. Other studies that investigate the impact of this pervasive narrative on decisions of promotion, retention, and funding would be particularly valuable. Expectations, whether perceived or enforced through tenure decisions, might give rise to some of our results. If these expectations vary from field to field, it is possible that while diversity remains a feature that spans academia, some types of trajectories may be more common in certain fields. Regardless of whether this is borne out by studies of other fields, models of faculty productivity will need to be revisited and revised.

\section*{acknowledgements}
The authors thank Mirta Galesic and Johan Ugander for helpful conversations. All authors were supported by NSF award SMA~1633747; DBL was also supported by the Santa Fe Institute Omidyar Fellowship.

\bibliographystyle{apsrev4-1}
\bibliography{manual_refs} 


\pagebreak
\begin{center}
\textbf{\large Supporting Information}
\end{center}

\setcounter{equation}{0}
\setcounter{section}{0}
\setcounter{subsection}{0}
\setcounter{figure}{0}
\setcounter{table}{0}
\setcounter{page}{1}
\makeatletter
\renewcommand{\theequation}{S\arabic{equation}}
\renewcommand{\thefigure}{S\arabic{figure}}
\renewcommand{\bibnumfmt}[1]{[S#1]}
\renewcommand{\citenumfont}[1]{S#1}

\section{\label{supp:cv_data}Collection of CV data}

Performing the DBLP coverage analysis and adjustment discussed in Supplemental Text~\ref{supp:data} required a benchmark dataset with complete coverage of the publications histories of a representative subset of researchers. This Supplemental Text describes the relevant details of the collection of that benchmark dataset. We manually extracted lists of publication dates from publicly available curricula vitae (CVs) belonging to a  random 10\% of the $N\!=\!1091$ researchers with career lengths between 10 and 25 years and having publications in at least three distinct years. This last condition ensures that the piecewise linear model can be fit to the individual's trajectory and excludes just 32 researchers from our analysis. Because of the high diversity of productivity trajectories, we chose 10\% of individuals, uniformly at random, from each of the quadrants designated by the signs of the two slope parameters, $m_1$ and $m_2$, as shown in Figure~\ref{fig:piecewise_shapes}.
Specifically, names of researchers from each quadrant were randomly shuffled and then collected, in order, until reaching 10\% of the total. However, individuals for whom a CV could not be found, or whose publicly available CV was last updated before 2011 were skipped, and other faculty were randomly selected in their place. Success rates for this exercise ranged between 66.3\% and 86.6\%, measured as the number of successfully extracted publication lists versus the total number of attempts. The majority of these failures were due to researchers having out-of-date CVs. Future studies should consider whether such partial records of researcher productivity are sufficient for analysis, as their inclusion would greatly improve success rates during collection.

\section{\label{supp:data}General trends in productivity data}

Meaningful trends in publication rates over a career can be confidently identified from raw publication counts only if two conditions are met. First, raw publication counts must be exhaustive, containing all peer-reviewed publications. Second, field-wide publication rates must be stationary over time. Due to the facts that DBLP data do not satisfy the former, and that computer science as a field does not satisfy the latter, raw publication counts recorded in the DBLP dataset must be adjusted to compensate before they can be analyzed. This Supplemental Text explains the details of two compensatory adjustments to DBLP data. We justify the first adjustment by providing a detailed analysis of time-varying fraction of publications covered by the DBLP dataset, anchored by a hand-collected benchmark CV dataset (see Supplemental Text~\ref{supp:cv_data}). We then justify the second adjustment by identifying a clear and significant overall growth in publication rates over forty years of computer science publication data. We conclude by discussing several possible explanations for why researcher productivity increases over time.

DBLP has indexed the overwhelming majority of current computer science publications, including peer-reviewed conferences and journals. However, while excellent today, this coverage has increased systematically over time, meaning that DBLP coverage is less complete for older publications and faculty. In order to quantify trends in the time-varying coverage of our DBLP data, we hand-collected the CVs of 109 faculty, representing 10\% of individuals whose trajectories were shown in Figure.~\ref{fig:piecewise_shapes},
providing a set of benchmark publication lists (see Supplemental Text~\ref{supp:cv_data}).

\begin{figure}[h!]
	\centering
	\includegraphics[width=0.48\textwidth]{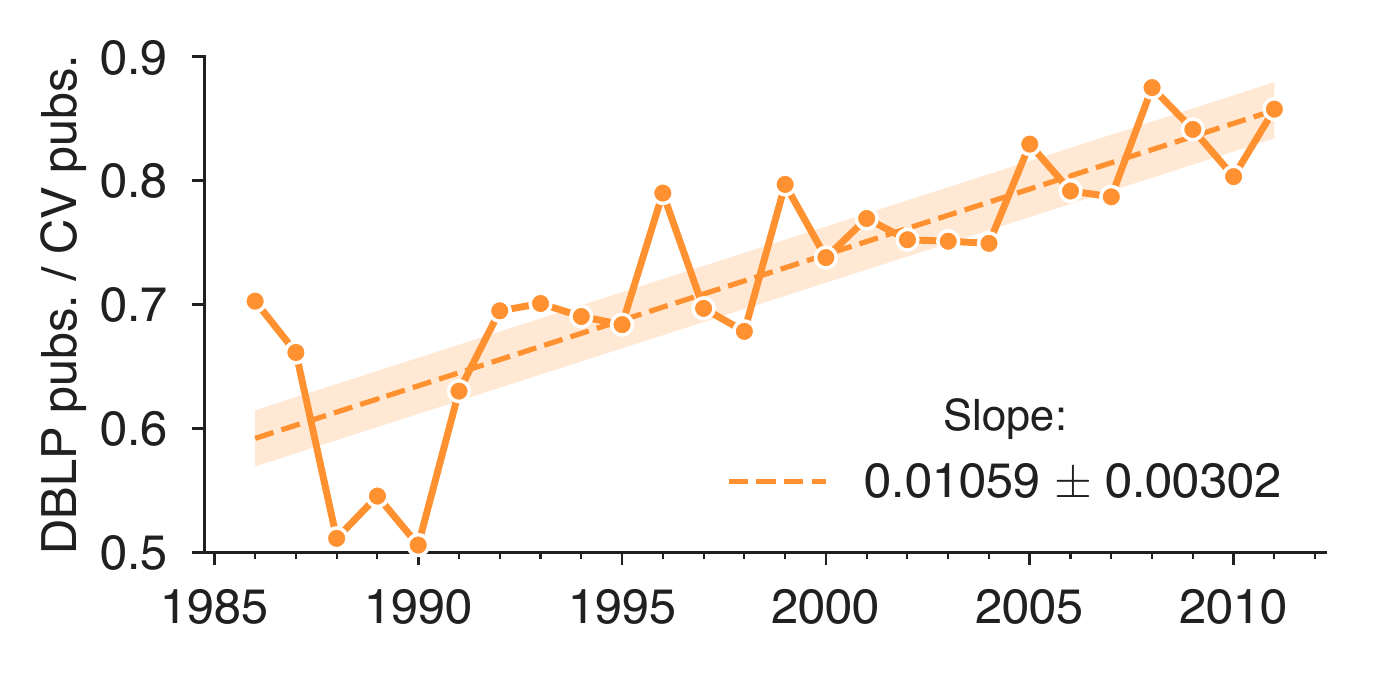}
	\caption{{\bf DBLP coverage improves for more recent publications.} {\normalfont Fraction of all publications found in DBLP data compared to publication lists extracted from CVs of corresponding researchers, separated by year. Regression of these fractions reveals that DBLP coverage improves by approximately 1.06\% each year. Shaded region denotes the 95\% confidence interval for the regression.}}
	\label{fig:dblp_data}
\end{figure}

For each year of our dataset, we compared the number of publications listed on individuals' CVs to the number of publications listed on their corresponding DBLP profiles, selecting only peer-reviewed conference and journal publications from CVs. Comparing DBLP data with CV benchmarks two sets of counts reveals that DBLP coverage has increased linearly from around 55\% in the 1980s to over 85\% in 2011 (Figure~\ref{fig:dblp_data}). DBLP coverage has grown at a rate of approximately 1.06\% of additional coverage per year, with a 95\% confidence interval indicating that this rate falls between 0.8\% and 1.4\%. Because the ratio of DBLP publications $y_{\text{DBLP}}$ to CV publications $y_{\text{CV}}$ is well described by the line
\begin{equation}
	\frac{y_{\text{DBLP}}(t)}{y_{\text{CV}}(t)} = m_\alpha t + b_\alpha\ ,
	\label{adj_dblp}
\end{equation}
we use Eq.~\eqref{adj_dblp} to convert all non-benchmarked DBLP publication counts to CV-equivalent publication counts, with estimated parameters of $\hat{m}_\alpha = 0.010588$ and $\hat{b}_\alpha = -20.434804$. 

After linearly adjusting all raw publication counts to correct for the expected DBLP coverage in a given year [Eq.~\eqref{adj_dblp}], we analyzed how individual researcher productivity has changed over the years spanned by our dataset. Due to the fact that we extracted and adjusted DBLP records for only authors in the faculty hiring dataset \cite{clauset2015systematic,way2016gender}, a straightforward analysis of the number of per-person publications in each calendar year would feature a different mixture of career ages in each year. For example, adjusted publication counts from the 1970s would include only early-career researchers; late-career researchers in the 1970s retired long before our dataset was collected. Because the main text of this paper reveals systematic trends in productivity by career age, this straightforward counting technique would introduce bias.

\begin{figure}[h!]
	\centering
	\includegraphics[width=0.475\textwidth]{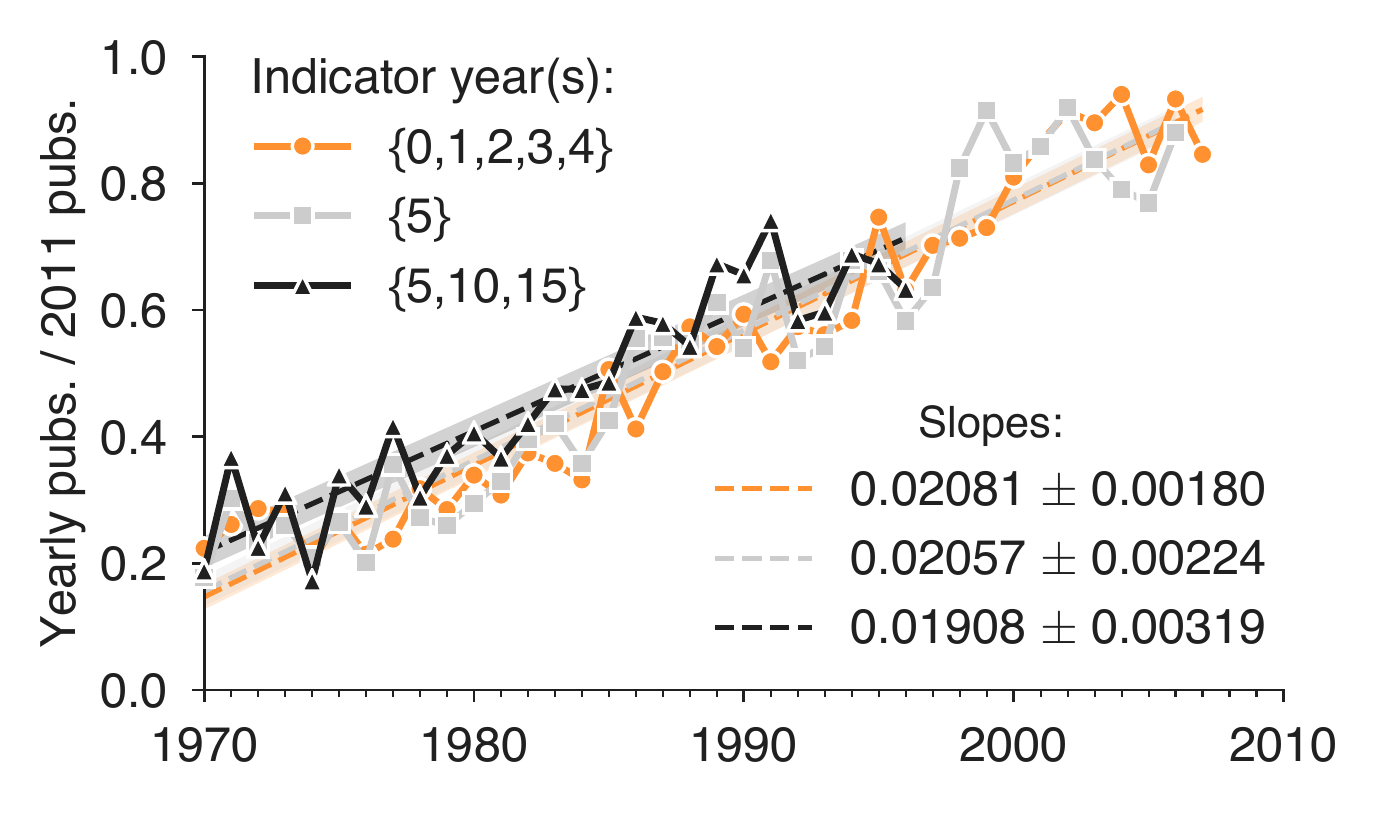}
	\caption{{\bf Individual productivity has increased over time.} {\normalfont After adjusting for DBLP coverage (see Supplemental Text~\ref{supp:data}), evaluation of average individual performances in select indicator years reveals that researchers have become more productive over time, growing at a rate of approximately one additional paper per decade. Shaded regions denote the 95\% confidence interval for each regression.}}
	\label{fig:supp_pub_growth}
\end{figure}

To quantify the expansion of publication rates over time, without introducing career-age bias, we selected ``indicator'' career ages at which to measure productivity, and compared how productivity at specific points in the academic career has changed over time. The trend in publication growth is consistently positive for all indicator career ages, and as in the main text, publication rates after year 5 tend to be higher than publication rates in the first four years. When these publication rates were normalized by their 2011 values, all indicator sets collapsed onto a common growth line (Figure~\ref{fig:supp_pub_growth}). Thus, while the indicator sets of $\{0,1,2,3,4\}$, $\{5\}$, and $\{5,10,15\}$, revealed that productivity grows at rate between 0.84 and 1.48 additional papers per person per decade, their rates of growth are directly proportional to 2011 productivity, meaning that the shape of the canonical trajectory has not changed over time, but has simply expanded proportionally.

The relative slopes of publication rate expansion are consistent with the relative publication rates in the canonical ``average'' productivity trajectory (Figure~\ref{fig:prod_curves_by_rank}).
Indeed, growth in researchers' first five years of productivity is relatively modest, which is expected since researchers, both historically and more recently, spend these years building their research programs by applying for funding and recruiting graduate students and postdoctoral researchers. On the other hand, productivity in year five---the year of or immediately preceding tenure evaluations at most institutions and, perhaps not coincidentally, the modal year of peak researcher productivity---grows at a faster rate of 1.48 additional papers per person per decade. These rates are both similar comparing years 5, 10, and 15, which describes changes across a larger window of career productivity.

The relationship between 2011-equivalent publications and past publications is consistently linear over the time spanned by our dataset (Figure~\ref{fig:supp_pub_growth}) and is modeled well by
\begin{equation}
	\frac{y_{\text{CV}}(t)}{y_{\text{2011}}(t)} = m_\beta t + b_\beta .
	\label{adj_growth}
\end{equation}
We therefore used Eq.~\eqref{adj_growth} to convert all CV-equivalent publication counts to 2011-equivalent publication counts, with estimated parameters $\hat{m}_\beta = 0.131873$ and $\hat{b}_\beta = -258.286620$. Thus, we applied the two transformations of ~\eqref{adj_dblp} (Figure~\ref{fig:dblp_data}) and Eq.~\eqref{adj_growth} (Figure~\ref{fig:supp_pub_growth}) in series to publication data to produce the adjusted publication counts used in the main text, unless otherwise specified; benchmark CV data was used for the 109 individuals for whom we collected it, and for those individuals, only Eq.~\eqref{adj_growth} was applied. Note that while model-based correction can adjust for trends in publication rates, alternative model-free approaches have been used by other researchers which convert publication rates to annual publication rate $z$-scores, e.g. \cite{duch2012possible}.

\begin{figure}[h!]
	\centering
	\includegraphics[width=0.475\textwidth]{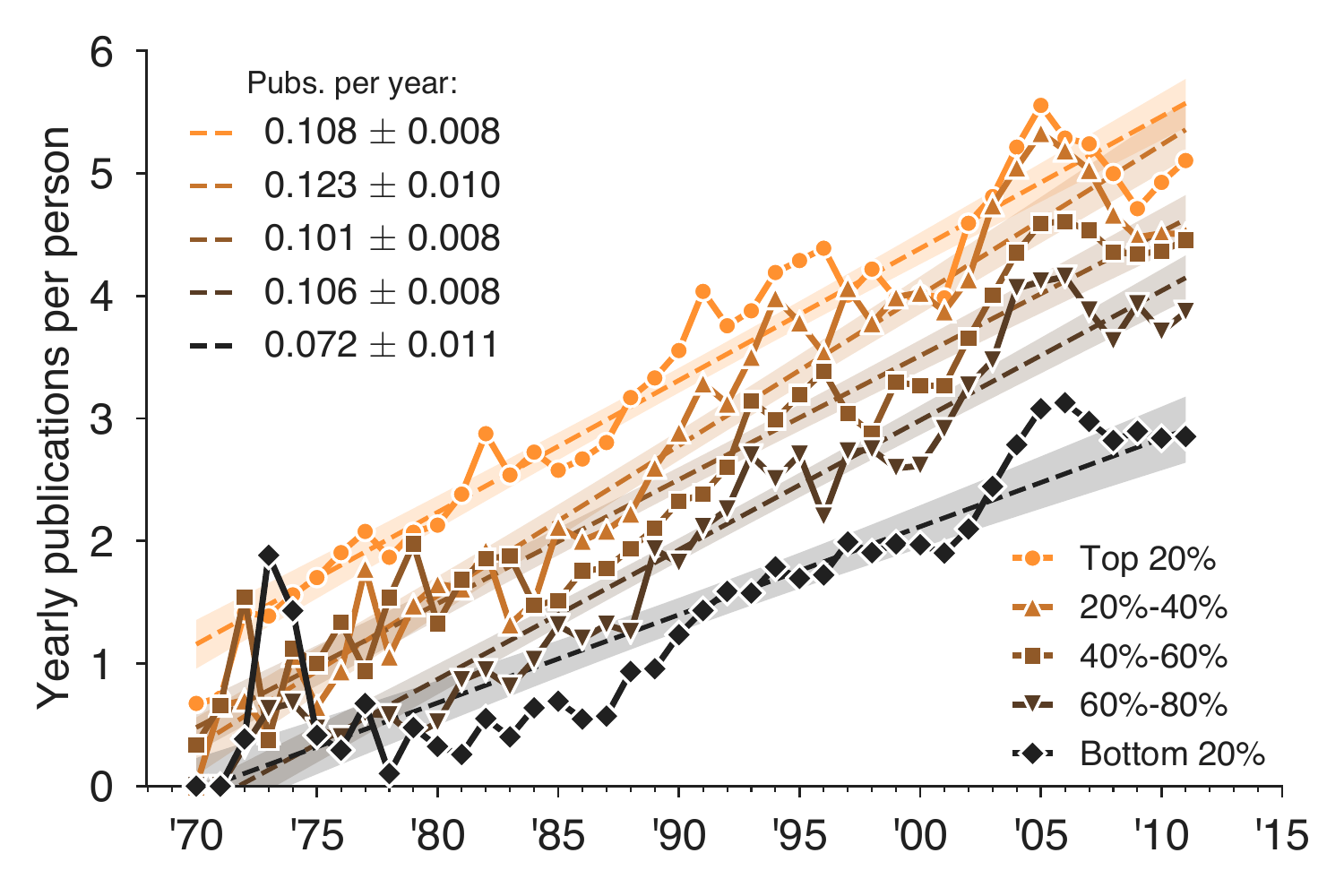}
	\caption{{\bf Annual publication rates have grown steadily.} {\normalfont For faculty in this study, per-person annual publications have increase over time at a rate of approximately one additional paper every 10 years. This rate of growth affects researchers at all levels of prestige rank $\pi$ \cite{clauset2015systematic}. Slopes represent least-squares linear regressions, with shaded regions denoting corresponding 95\% confidence intervals.}}
	\label{fig:growth}
\end{figure}

In the main text, we noted that, after applying these two linear adjustments, the median number of early-career publications per person per institution increases over time and correlates with prestige. Figure~\ref{fig:growth} illustrates this trend, stratifying individuals into three levels of prestige, and revealing that production growth rates between higher and lower prestige departments have widened slightly but significantly over time ($p\!<\!0.05$, two-tailed $t$-test).

\begin{figure}[h!]
	\centering
	\includegraphics[width=0.475\textwidth]{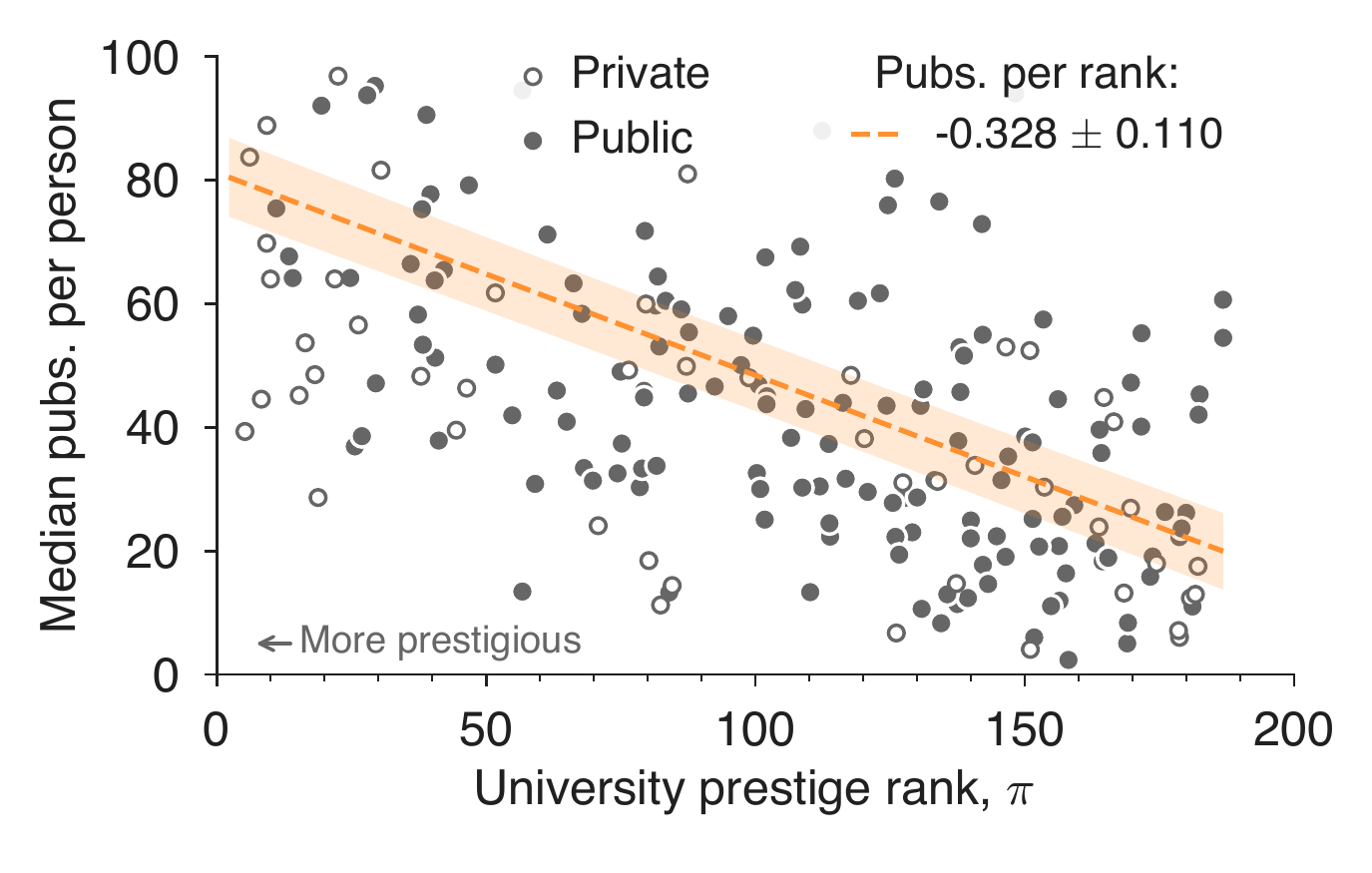}
	\caption{{\bf Publications correlate with prestige of employing institutions.} {\normalfont Dots indicate median number of publications per person per institution for all years post-hire, adjusted for growth in publication rates over time and ordered by institutional prestige. Effects of prestige are statistically indistinguishable for private (open dots) and public (closed dots) institutions. Shaded region denotes the 95\% confidence interval for least squares regression.}}
	\label{fig:supp_pub_priv}
\end{figure}

This trend, observed for early-career publications (i.e., publications within the first ten years of a career, for individuals with careers of ten years or longer) is no different from the trend for all post-hire publications and all researchers (Figure~\ref{fig:supp_pub_priv}). Median lifetime career productivity correlates significantly with prestige, and, as in early-career productivity, public and private institutions are similarly affected by this relationship. Using an ordinary least squares regression of productivity versus prestige including dummy and interaction terms for public/private status we found that the relationship between productivity and prestige is not significantly affected by public/private status ($p\!>\!0.05$, $t$-test, for both public/private dummy and public/private-prestige interaction).

\begin{figure}[t]
	\centering
	\includegraphics[width=0.8\linewidth]{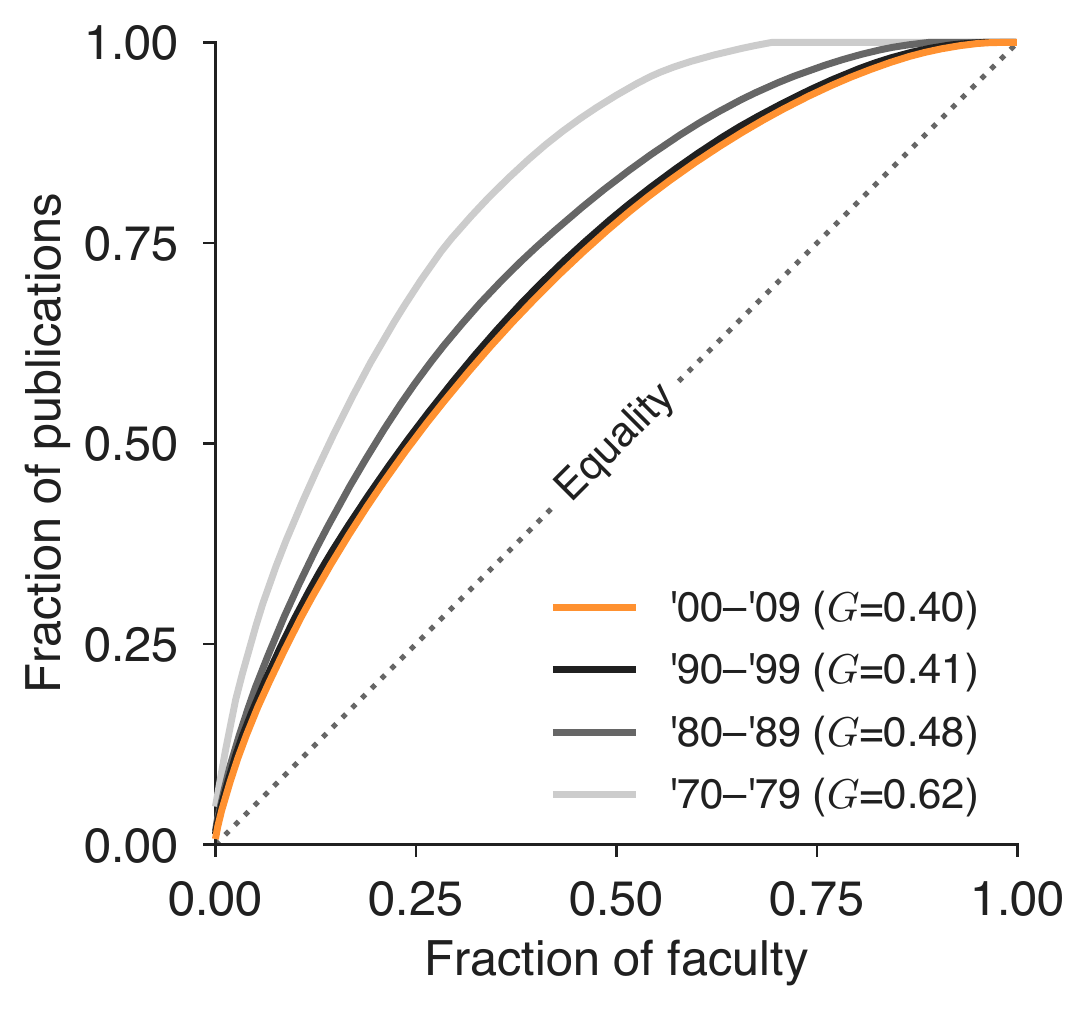}
	\caption{{\bf Production imbalance among faculty.} {\normalfont Lorenz curves of adjusted production by in-sample faculty, stratified by decade of first hire (see legend), show that approximately 20\% of faculty account for half of all publications in the dataset. The Gini coefficient $G$ for each curve is noted in the legend; diagonal line indicates equal production by all faculty.}}
	\label{fig:imbalance}
\end{figure}

While production rates have increased steadily over time and for all levels of prestige, we find that the imbalance in research production has decreased in recent years. Figure~\ref{fig:imbalance} illustrates this inequality across researchers and time by showing the fraction $Y$ of all publications in our sample that were produced by the most productive fraction $X$ of all faculty (a Lorenz curve), for faculty first hired in each of the four decades that our data span and restricting analysis to only publications produced in the first five years of an individual's career. As referenced in the main text, the Gini coefficients for productivity imbalance have declined, from 0.62 in the 1970s to 0.40 in the 2000s. Many factors could potentially drive such a shift towards more balanced research production in science (including, for example, technological advancements and corresponding declines in research costs that may have leveled the playing field for researchers at less-prestigious universities), and we welcome future studies that explore this shift in more detail.

Finally, several possibilities exist that might explain why researchers are becoming more productive. First, the average number of co-authors per publication has steadily increased over time, allowing researchers to work on a larger number of projects. Second, the number of publication venues has also grown, providing more outlets for researchers' work and potentially facilitating more specialized communities with faster peer review. Third, technological advances including improvements in computer architecture have benefitted researchers universally, increasing the speed at which results are both generated and published. Lastly, perhaps the perceptions of what constitutes the minimum publishable unit of research has changed over time, resulting in a larger number of shorter, more narrowly focused publications in recent years.

\section{\label{supp:modeling}Modeling framework}

Equation~\eqref{eqn:piecewise} is the simple model used in the main text to parameterize the adjusted publication counts over the course of a career. Reproduced below, it consists of two lines with slopes $m_1$ and $m_2$ that intersect at time $t^*$. 
\begin{equation}
f(t) = 
\begin{cases} 
      b+m_1 t & 0\leq t\leq t^* \\
      b+m_1 t^* + m_2 (t-t^*) & t>t^*
\end{cases}
\label{eqn:piecewise_sup}
\end{equation}
In this manuscript, we fit Eq.~\eqref{eqn:piecewise_sup} to adjusted count data using least squares. However, there exist other regression frameworks that correspond to generative models for time series data. In this section, we discuss some of these alternatives.

\subsection{Adjusting models instead of counts}

Although publication time series are naturally count data, a regression framework that is naturally suited for counts, such as Poisson or Negative-Binomial regression, is not advised. Directly fitting raw counts using a Poisson or Negative-Binomial model would neglect the adjustments for both the coverage of DBLP and the time-varying changes in publication rates (Supplemental Text \ref{supp:data}). On the other hand, adjusted publications are non-integers, rendering them inappropriate for count regressions. However, there are alternatives that would allow for both count regressions and the adjustments of Supplemental Text \ref{supp:data}. These adjustments come with a price, however, due to the assumptions and free parameters that they introduce. 

One alternative solution to fitting a model to adjusted publication data is to fit an adjusted linear model to raw publication data. In other words, adjust the model instead of the data. Due to the fact that this approach would preserve the data as counts, this model would be amenable to Poisson and Negative-Binomial regression frameworks. Adjusted publications $y_{\text{2011}}$ are related to raw publications by the adjustment
\begin{equation}
	y_{\text{2011}}(t) = y_{\text{DBLP}}(t) \frac{1}{\hat{m}_{\alpha}t + \hat{b}_{\alpha}} \frac{1}{\hat{m}_{\beta} t + \hat{b}_{\beta}}
\end{equation}
where $m_\alpha$ and $m_\beta$ are slopes and $b_\alpha$ and $b_\beta$ are intercepts of the linear adjustments for DBLP coverage and publication expansion, respectively, and hats indicate that variables have been estimated from data (Supplemental Text \ref{supp:data}). Applying this adjustment to the model $f$, which is to hold for adjusted publications, we get
\begin{equation}
	\label{seqn}
f_{\text{DBLP}}(t) = (\hat{m}_{\alpha}t + \hat{b}_{\alpha})(\hat{m}_{\beta}t + \hat{b}_{\beta}) f(t-t_0)
\end{equation}
where $t_0$ is the initial year of a particular faculty member's career, and $t$ is the calendar year. 

We now turn to details of Poisson and Negative-Binomial frameworks for fitting the adjusted model Eq.~\eqref{seqn}, and discuss the assumptions and parameters that they introduce.

\subsection{Poisson Model}

Consider a Poisson fit of Eq.~\eqref{seqn} to a set of data given by $\{t_i, y_i\}$. To simplify, let us be explicit about the dependence of $f_{\text{DBLP}}$ on the four parameters, $m_1$, $m_2$, $b$, and $t^*$, which we collectively refer to as $\theta$.
$$f_{\text{DBLP}}(t;\theta) = q(t) f(t-t_0;\theta)$$
where we have made clear that $q(t) = (\hat{m}_{\alpha}t + \hat{b}_{\alpha})(\hat{m}_{\beta}t + \hat{b}_{\beta})$ does not depend on the model parameters $\theta$. The likelihood is then
\begin{equation}
	P(\{t_i, y_i\} | \theta) = \prod_{i} \frac{e^{-q(t_i)f(t_i-t_0;\theta)}\left [q(t_i)f(t_i-t_0;\theta)\right ]^{y_i}}{y_i !}
\end{equation}
Rather than maximizing $P$, we will maximize $\log P$. Taking the natural log of both sides, we get
\begin{align}
	\log P(\{t_i, y_i\} | \theta) &= \sum_{i} \bigg \{ -q(t_i)f(t_i-t_0;\theta) \nonumber \\ 
	&\quad + y_i \left [\log q(t_i) + \log f(t_i-t_0;\theta) \right] - \log{y_i !} \bigg \}
\end{align}
Note that the terms $-\log{y_i !}$ and $y_i \log q(t_i)$ do not depend on the parameters $\theta$, so they affect the value of the maximum but not its location in parameter space. Dropping them yields a log-likelihood score $\mathcal{L}$ of
\begin{equation}
	\mathcal{L}(\{t_i, y_i\} | \theta)  = \sum_{i} -q(t_i)f(t_i-t_0;\theta) + y_i \log f(t_i-t_0;\theta)\ .
	\label{poissonfit}
\end{equation}

Note that for any trajectory, $q(t_i)$ can be precomputed and does not depend on the parameters $\theta$. Thus, fitting the 2011-equivalent Poisson model requires that we maximize Eq.~\eqref{poissonfit} with respect to $m_1$, $m_2$, $b$, and $t^*$. This equation must be maximized numerically.

While this adjusted Poisson model is attractive because it naturally fits count data, it imposes assumptions on the data-generating process that are not justified empirically. Namely, the variance and mean of a Poisson distribution are equal, meaning that the Poisson regression expects the same of the data it explains.

\subsection{Negative Binomial Model} The Poisson model above enforces the constraint that the mean is equal to the variance. However, there is no indication that the data support this assumption so we introduce the standard alternative, the Negative Binomial model. This model requires both a mean $\mu$ and a heterogeneity parameter $\zeta$, such that the probability of a single observation $y$ is
\begin{equation}
	P(y) = \frac{\Gamma(y+\frac{1}{\zeta})}{\Gamma(y+1)\Gamma\left(\frac{1}{\zeta}\right)} \left ( \frac{1}{1+\zeta \mu} \right )^{\frac{1}{\zeta}} \left ( \frac{\zeta \mu}{1+\zeta \mu} \right )^{y}
\end{equation} 

As in the Poisson regression, we will once more parameterize the mean using the piecewise linear model as $\mu_i = \mu(t_i) = q(t_i)f(t_i - t_0)$. However, we must also introduce a model for $\zeta(t_i)$. 

The easiest way forward, mathematically, is to set $\zeta(t_i) = \zeta$. This assumes equal heterogeneity around the expected value $\mu(t_i)$ for all time points in a career $t_i$. Note that this assumption decouples the heterogeneity from the mean, while under the Poisson model they are directly coupled. One might think of the Poisson model therefore as fitting the parameters $\theta$ to both the trend and the fluctuations together. The fixed-$\zeta$ Negative Binomial model, on the other hand, fits the parameters $\theta$ to the trend and uses a fixed $\zeta$ to accommodate all fluctuations. In this sense this Negative Binomial approach is more flexible, and uses an additional parameter to gain that flexibility. 

The $\zeta(t_i) = \zeta$ assumption results in a log probability of 
\begin{align}
	\log P(\{t_i, y_i\} | \theta,\zeta) = \sum_{i=1}^{T} \bigg \{ \log \Gamma\left(y_i+\frac{1}{\zeta}\right) - \log \Gamma(y_i+1)  \nonumber \\ 
	 - \log \Gamma\left(\frac{1}{\zeta}\right) - \frac{1}{\zeta} \log \big[1+\zeta q(t_i)f(t_i - t_0)\big] \nonumber \\
	  + y_i \big ( \log \big[\zeta q(t_i)f(t_i-t_0;\theta)\big] - \log  \big[1+\zeta q(t_i)f(t_i - t_0;\theta)\big]\   \big )\bigg \}
\end{align}
and we note that $\sum_{i=1}^{T} \log \Gamma\left(\frac{1}{\zeta}\right) = T \log \Gamma\left(\frac{1}{\zeta}\right)$, and that both $\sum_{i=1}^{T} \log \Gamma(y_i+1)$ and $\sum_{i=1}^{T} y_i \log q(t_i)$ are constants that do not depend on either $\theta$ or $\zeta$, allowing us to write a log-likelihood score of 
\begin{align}
	\mathcal{L}(\{t_i, y_i\} | \theta,\zeta) = - T \log \Gamma\left(\frac{1}{\zeta}\right) + \sum_{i=1}^{T} \bigg \{ \log \Gamma\left(y_i+\frac{1}{\zeta}\right) \nonumber \\ 
	- \frac{1}{\zeta} \log \big [1+\zeta q(t_i)f(t_i - t_0) \big] \nonumber \\
	+ y_i \big ( \log \big[\zeta f(t_i-t_0;\theta)\big] - \log  \big[1+\zeta q(t_i)f(t_i - t_0;\theta)\big]\ \big )\bigg \}
\end{align}
Progress here, however, is obstructed by the difficulties of taking derivatives of Gamma functions. Thus, the above equation must be optimized numerically over the parameters $\theta$ and $\zeta$. 

While these calculations may be helpful in seeding a way forward in future work, it is important to note that the fixed-$\zeta$ Negative Binomial model also makes strong assumptions about the generative process that created the data. Indeed, the assumption that fluctuations are uniform over an entire career is strong, and is not justified  by data. One could also avoid this assumption, but this introduces additional problems, which we now discuss.

The temptation to let each point $t_i$ have a parameterized value of $\mu(t_i)$ and a free parameter of $\zeta(t_i)$ results in overfitting. Note that this would allow each point in the time series $(t_i,y_i)$ to be fit by a negative binomial distribution with a mean given by Eq.~\eqref{seqn} and an arbitrarily large or small $\zeta(t_i)$, resulting in dramatic overfitting. This approach therefore makes few assumptions, but provides little value to the modeler. 

A middle ground between fixed $\zeta$ and unrestricted $\zeta(t_i)$ would be to parameterize $\zeta(t_i)$ using a lower dimensional model. Using the same model for $\zeta(t_i)$ as we used for $\mu(t_i)$ would be similar, in principle, to the Poisson regression. Using a different model for $\zeta(t_i)$ is an option, but would require, again, a deep focus on the underlying mechanisms hypothesized to explain fluctuations in productivity. 

\subsection{Modeling outlook}

Generative models for productivity trajectories would be enormously valuable. In this Supplemental Text, we emphasized the assumptions made by the generative models underlying various regression frameworks. In particular, we derived models that are able to be fit directly to raw count data by including the inverse of the adjustment derived in Supplemental Text~\ref{supp:data}, a quadratic term referred to as $q(t)$. 

In terms of impacts, fitting the Poisson model and the fixed-$\zeta$ Negative Binomial model to the trajectories investigated in this paper do affect the parameters of individuals' trajectories. However, they do not diminish the diversity of trajectories that we observe. Indeed, the example trajectories shown in Fig.~\ref{fig:piecewise_shapes} are only subtly affected by the use of one type of generative model or another.


\begin{figure}[t]
	\centering
	\includegraphics[width=0.475\textwidth]{\figfolder 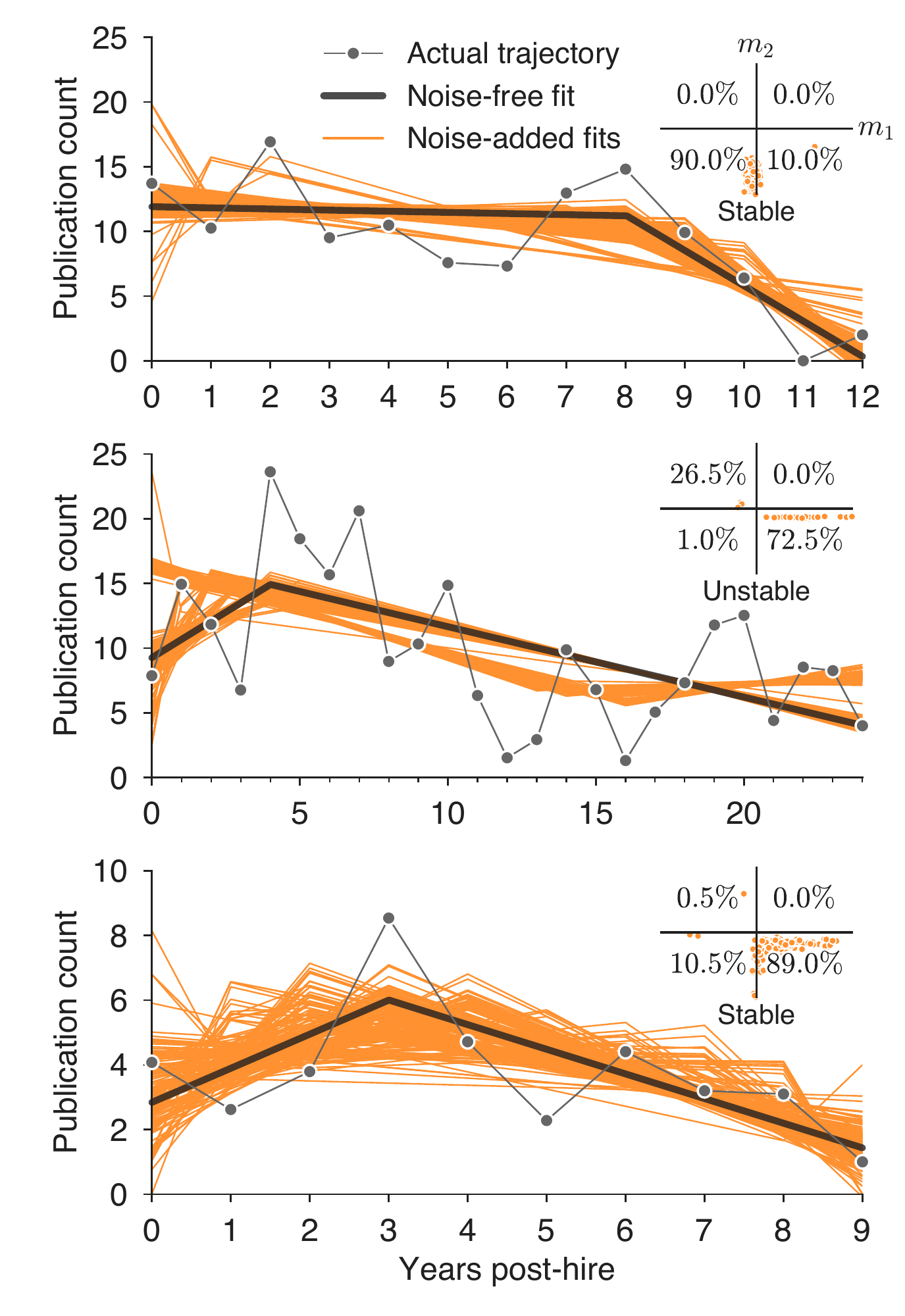}
	\caption{{\bf Example model fits for noise-free and noise-added publication trajectories.} {\normalfont Example publication trajectories (gray, dotted) with piecewise linear fits (black) and 200 fits to noise-added trajectories (orange). Trajectories are categorized as stable whenever 75\% or more of fits fall within a single quadrant, as indicated in inset diagrams.}}
	\label{fig:noisy_fits}
\end{figure}

\section{\label{supp:sensitivity}Sensitivity to timing of publications}

Publication generally signifies the conclusion of a research project but the exact date when an article is published can depend on many factors, including the availability of reviewers, graduation deadlines for graduate students, delays between acceptance and publishing, synchronization with conference submission deadlines, as well as non-academic constraints, such as the impending birth of a child. Each of these factors might advance or delay a publication's appearance in the literature, and furthermore, the effort associated with each publication may span weeks, months, or years. As a result, publication years serve as a noisy indicator of when productivity occurs.

\begin{figure}[t]
	\centering
	\includegraphics[width=0.475\textwidth]{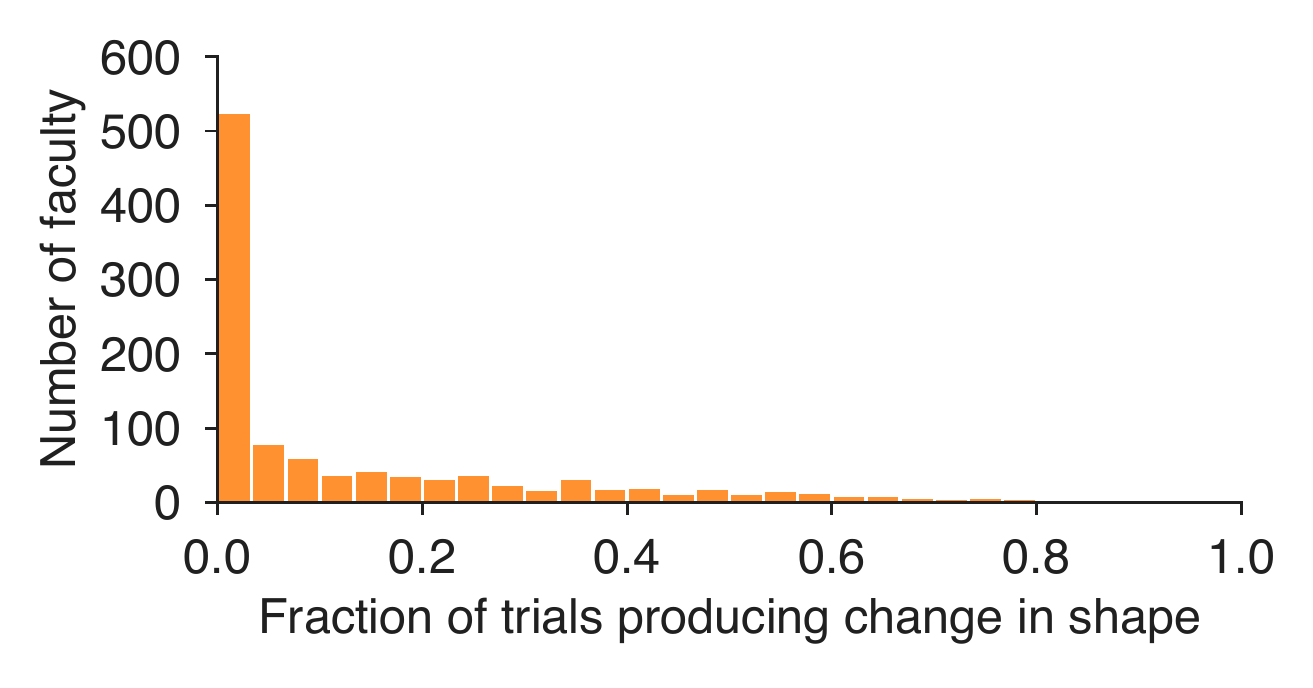}
	\caption{{\bf Trajectory shapes are robust to perturbations in publication years. }{\normalfont Applying the piecewise linear model to 200 noise-added publication trajectories for each researcher, the median fraction of trials resulting in sign changes of model parameters $m_1$ or $m_2$ compared to noise-free fits is 0.045.}}
	\label{fig:noisy_fracs}
\end{figure}

To ensure that our findings are not due to coincidence in the timings of researchers' publications, we examined the sensitivity of our results to the addition of small amounts of noise. For each researcher and for each of their publications, we added noise drawn from a normal distribution ($\mu\!=\!0$, $\sigma\!=\!0.7413011$) to the publication year and then rounded to the nearest whole-year. In expectation, this process leaves one half of publication years unaffected and shifts 22.8\% by one year in either direction, 2.1\% by two years, and 0.04\% by three years. We repeated this process 200 times for each researcher (examples shown in Figure~\ref{fig:noisy_fits}), and found that the median number of trials in which the trajectory changed shape compared to the noise-free fit (i.e., $m_1$ or $m_2$ changed sign) was 9 (4.5\%; see Figure~\ref{fig:noisy_fracs}).

While the typical individual's parameters are robust to noise, those individuals whose trajectories featured few publications or whose noise-free model parameters were near zero were far more likely to change shape. In fact, for 10.5\% of individuals, noise led to shape change more often than not, i.e in greater than 50\% of noisy repetitions. Therefore, as an additional check, we asked whether the model parameters inferred for researchers' noise-free trajectories differ significantly from those inferred for their 200 noise-added trajectories. We used Fisher's method to combine $p$-values for each researcher and found no significant differences in any of the four model parameters ($m_1$, $m_2$, $b$, and $t^*$). Evaluated separately, fewer than one percent of researchers' inferred model parameters differ significantly under the noise-free and noise-added fits. We conclude that the general shape of productivity trajectories is robust to small differences in publication year. 

\begin{figure}[t]
	\centering
	\includegraphics[width=0.45\textwidth]{\figfolder 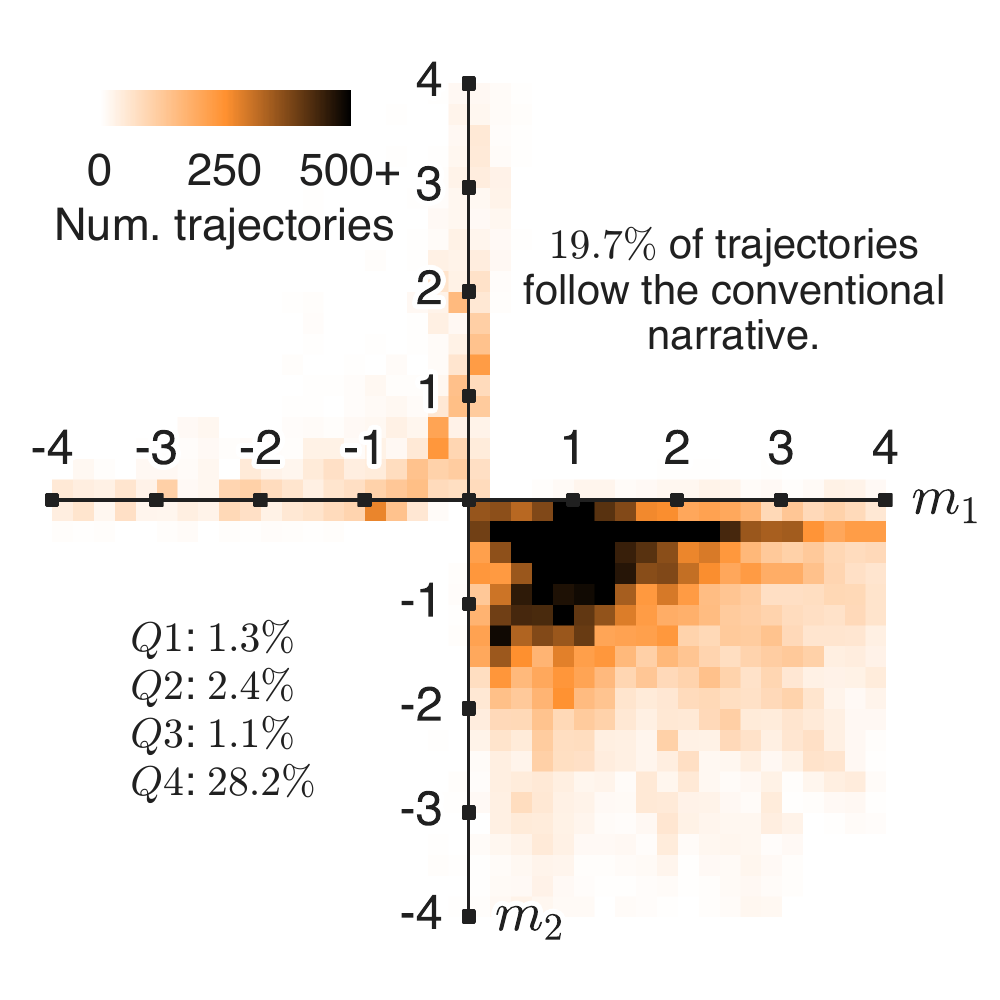}
	\caption{{\bf The distribution of noise-added trajectories matches that of individuals.} {\normalfont The distribution of slope parameters $m_1$ and $m_2$ for 200 noise-added trajectories, for each individual and requiring that each instance is not better modeled as a straight line (AIC with finite-size correction). Counts of noise-added trajectories are tabulated as percentages falling into each quadrant and within the octant corresponding to the conventional narrative (19.7\%).}}
	\label{fig:noisy_kde}
\end{figure}

Having specified a model for noise around each publication date, an individual's trajectory naturally becomes a distribution of trajectories, which in turn maps into a corresponding distribution in the $m_1\times m_2$ parameter space. To analyze this distribution, we inferred model parameters for each of a researcher's 200 noise-added trajectories, and for those trajectories that were more confidently modeled by Eq.~\eqref{eqn:piecewise} (using AIC with finite-size correction; see Supplemental Text~\ref{supp:model_selection}) versus a straight line, we compiled the distribution depicted in Figure~\ref{fig:noisy_kde}.

This distribution suggests a complementary approach to investigating the universality of the conventional narrative wherein individuals are considered as distributions rather than point estimates. The latter of these approaches (presented in the main text) requires specifying a threshold for stability that determines whether or not an individual can be confidently mapped into a particular region in the space. The former, on the other hand, requires no such distinction and instead sums the independent distributions of individual faculty to quantify the total probability mass in each region. The point estimate approach makes a statement about individuals, while the distribution approach makes a statement about the population. Importantly, both statements agree: approximately 20\% of individuals are firmly mapped to the canonical octant, and approximately 20\% of probability mass maps into the canonical octant.

\section{\label{supp:model_selection}Model selection}

When the complexity of a model exceeds the complexity of the underlying data, some parameters of the model may no longer be interpreted as meaningful. Although the piecewise linear model of Eq.~\eqref{eqn:piecewise} has only four parameters---two slopes, $m_1$ and $m_2$, an intercept $b$, and a change point $t^*$---it may nevertheless overfit productivity trajectories that are actually linear. This Supplemental Text provides additional details for our model selection procedures that avoid the overinterpretation of the piecewise change point parameters $t^*$.

If a publication trajectory is generated by a straight line with added noise, then fitting a piecewise model will result in  $m_1$ approximately equal to $m_2$, and the location of the change point $t^*$ will be arbitrary. We apply model selection to identify individuals with careers lengths between 10 and 25 years ($N\!=\!1091$) whose productivity trajectories are both stable (see Supplemental Text~\ref{supp:sensitivity}) and consistently better modeled by the piecewise model Eq.~\eqref{eqn:piecewise} than a straight line (i.e., ordinary least squares or OLS). This filter includes only those trajectories whose change points $t^*$ can be interpreted with confidence.

To perform model selection, we consider three information theoretic model selection techniques: Akaike information criterion (AIC), Akaike information criterion with finite-size correction (AIC$_{\text{c}}$), and Bayesian information criterion (BIC), defined as
\begin{align*}
\text{AIC} &: n\log(SSE/n) + 2k \\
\text{AIC}_{\text{c}} &: n\log(SSE/n) + 2k + {2k(k\!+\!1)\over{n\!-\!k\!-\!1}} \\
\text{BIC} &: n\log(SSE/n) + k\log(n),
\end{align*}
where $n$ is length of the individual's career, $k$ is the number of model parameters [$k\!=\!2$ for OLS, and $k\!=\!4$ for Eq.~\eqref{eqn:piecewise}], 
and $SSE$ is the sum of squared errors (differences) between actual and modeled publication counts. AIC is the least conservative of the three methods, finding that 59.1\% ($N\!=\!645$) of individuals are better modeled by Eq.~\eqref{eqn:piecewise}
 than OLS regression. By contrast, AIC$_{\text{c}}$ and BIC select only 33.2\% ($N\!=\!363$) and 44.4\% ($N\!=\!485$) of individuals, respectively. In the main text, we adopt the most conservative approach, AIC$_{\text{c}}$, but the other two methods nevertheless produce qualitatively similar distributions for $t^*$, with the modal year for $t^*$ remaining at year 5. 

\section{\label{supp:alphabetized}Detection of alphabetized publication venues}

Conventions of author order vary widely in computer science. In the first/last convention, first authorship is reserved for the lead author or primary contributor to the study, while last authorship indicates the senior author who oversaw or advised the work. In the alphabetical convention, borrowed from mathematics, a paper's authors are arranged alphabetically by last name. If the trend from first-authorship toward last-authorship over a career is to be reliably interpreted (Figure~8), publications with alphabetical author orders must be discarded. This Supplemental Text explains the methods used to statistically identify and remove publication venues that are highly enriched with alphabetical conventions.

Our approach is to count the number of multi-author papers in each publication venue with alphabetically ordered authors and compare this count to the number expected by chance. (We note that all single-author papers are ignored in this analysis.) A paper with $M$ authors will list its authors alphabetically by chance with probability ${1/{M!}}$. Noting the number of authors on each multi-author paper published by a particular venue and assuming independence of ordering decisions, we derive an empirical distribution representing the number of coincidentally-alphabetized author lists and ask whether the venue adopts the convention significantly more often than would be expected by chance. Additionally, we require that the number of observed alphabetized lists be at least twice the expected value. These two conditions ensure that the alphabetical convention is both significant and widespread in its adoption in a particular venue. We find that 630 of the 5622 (11.2\%) distinct venues in our dataset alphabetize their author lists. These venues account for 27,237 of the 177,437 (15.4\%) multi-author conference or journal publications for which the publication venue is known. 

Manually inspecting the list of alphabetized venues reveals that popular theoretical venues like STOC (Symposium on Theory of Computing), FOCS (Foundations of Computer Science), STACS (Symposium on Theoretical Aspects of Computer Science), and SODA (Symposium on Discrete Algorithms) adhere to the alphabetical convention, while WWW (World Wide Web Conference), CSCW (Conference on Computer-Supported Cooperative Work and Social Computing), KDD (Conference on Knowledge Discovery and Data Mining), CHI (Conference on Human Factors in Computing Systems), and AAAI (AAAI Conference on Artificial Intelligence) do not, matching our expectations.

\section{\label{supp:leastsquares}Least squares fit of $f(t)$}

A least-squares fit of the continuous piecewise-linear equation $f(t)$ given in Eq.~\eqref{eqn:piecewise} involves minimizing the sum of squared errors over four parameters, $m_1$, $m_2$, $b$, and $t^{*}$. In this Supplemental Text, we provide details that make this fitting process rapid and maximally accurate. 

The model fit consists of two steps. First, we assume that $t^{*}$ is fixed, and find the optimal values of $m_1$, $m_2$, and $b$. In the second step, we search for the $t{*}$ whose corresponding optimal parameters provide the best fit. The sum of squared error $\varepsilon$ is given by
\begin{align}
	\varepsilon = &\frac{1}{2} \sum \left ( m_1 t_i + b - y_i \right )^2 \nonumber \\
	+ &\frac{1}{2}\sum' \left( m_1 t^{*} + m_2(t_i - t^{*}) + b - y_i \right) ^2
	\label{sse}
\end{align}
where $t^{*}$ is the change point, $\sum$ denotes the sum for all $t_i < t^{*}$, and $\sum '$ denotes the sum for all $t_i \geq t^{*}$. 

In the first step, we imagine $t^{*}$ to be fixed and simply take partial derivatives with respect to the three parameters, set each equal to zero, and solve. Setting $\nabla \varepsilon = 0$ yields three equations,

\begin{align}
	m_1 \left [ \sum t_i^2 + \sum' t^{*2} \right] + m_2 \sum ' t^{*} \left ( t_i - t^{*} \right) \nonumber \\ + b \left [ \sum t_i + \sum ' t^{*} \right ] = \sum y_i t_i + \sum' y_i t^{*}, \nonumber \\
	m_1 \sum' t^{*} \left ( t_i - t^{*} \right ) + m_2 \sum' \left (t_i - t^{*} \right)^2 \nonumber \\ + b \sum' \left (t_i - t^{*} \right) = \sum' y_i \left ( t_i - t^{*} \right), \nonumber \\
	m_1 \left [ \sum t_i + \sum ' t^{*} \right ] + m_2 \sum ' \left(t_i-t^{*}\right) \nonumber \\  + b \left [ \sum 1 + \sum' 1 \right] = \sum y_i + \sum' y_i\ .
	\label{linearsystem}
\end{align}

Thus, for a fixed value of $t^{*}$, the above equations provide a linear system of three equations for the three unknowns $m_1$, $m_2$, and $b$. The system can be solved numerically using any linear solver.

In the second step, we embed the optimization above within a search for the optimal $t^{*}$ value. For each proposed value of $t^{*}$, we use ~\eqref{linearsystem} to find the optimal $m_1$, $m_2$, and $b$ and then compute the associated error using ~\eqref{sse}, choosing the $t^{*}$ that minimizes $\varepsilon$. Initially, we propose a coarse grid at the level of $\Delta t^{*}= 0.1$, and then refine the grid by an order of magnitude locally around the best result, repeatedly, until the optimal $t^{*}$ is known to single precision. This procedure is fast, and due to the result of ~\eqref{linearsystem}, limits numerical search to a one-dimensional line.

\clearpage


\end{document}